\documentclass[12pt,a4paper,fleqn]{article}
\usepackage[textheight=23cm,textwidth=15cm]{geometry}
\usepackage{amsmath}
\usepackage{graphicx}
\usepackage[american]{babel}
\usepackage{here}
\usepackage[journal=jpcafh,articletitle=true,maxauthors=10,etalmode=truncate]{achemso}

\usepackage{lscape} 

\begin{document}

\begin{center}

{\LARGE\bf
 Resonance Effects in the Raman Optical Activity Spectrum of [Rh(en)\textsubscript{3}]\textsuperscript{3+}
}

\vspace{1cm}

{\large
Thomas Weymuth$^{a,}$\footnote{Corresponding author; e-mail: thomas.weymuth@phys.chem.ethz.ch; ORCID: 0000-0001-7102-7022}
}\\[4ex]

$^{a}$ Laboratory for Physical Chemistry, ETH Zurich, \\
8093 Zurich, Switzerland

September 30, 2019

\vspace{.41cm}

\end{center}

\begin{center}
\textbf{Abstract}
\end{center}
\vspace*{-.41cm}
{\small

Raman optical activity spectra of $\Lambda$-tris-(ethylenediamine)-rhodium(III) 
([Rh(en)\textsubscript{3}]\textsuperscript{3+}) have been calculated at 16 on-, near-,
and off-resonant wavelengths between 290\,nm and 800\,nm. The resulting
spectra are analyzed in detail with a focus on the observed resonance 
effects. Since several electronically excited states are involved, the
spectra are never monosignate, as is often observed in resonance Raman
optical activity spectra. Most normal modes are enhanced through these
resonance effects, but in several cases, de-enhancement effects are
found. The molecular origins of the Raman optical activity intensity for
selected normal modes are established by means of group coupling
matrices. In general, this methodology allows one to produce an intuitive 
explanation for the intensity behavior of a given normal mode. However,
due to the complex electronic structure of [Rh(en)\textsubscript{3}]\textsuperscript{3+},
there are some intriguing resonance effects the origins of which could
not be fully clarified in terms of group coupling effects.
Therefore, simple and general rules that
predict how the intensity of a specific normal mode is affected by
resonance effects are difficult to devise.

}

\newpage

\section{Introduction}
\label{sec:introduction}

Chiral molecules show a small difference in the intensities of Raman scattered left and right circularly polarized light, 
an effect generally denoted as Raman Optical Activity (ROA)\cite{Barron2004a}. This effect was predicted by Barron 
and Buckingham in 1971\cite{Barron1971} and first observed experimentally in 1973 by Barron \textit{et al.}\cite{Barron1973} 
(this observation was independently confirmed by Hug \textit{et al.}~two years later\cite{Hug1975}). Since the intensity 
difference measured in ROA spectroscopy is very small, it took more than twenty years and major improvements in instrumentation 
techniques\cite{Spencer1988, Hecht1994, Hug1999} before it was firmly established as the routine analytical method it 
is today\cite{Barron2000, Barron2004, Barron2007, Barron2010}. In particular, ROA spectroscopy is often adopted to study 
biomolecules in their natural, aqueous environment\cite{Barron2000}. However, ROA studies on chiral metal complexes are 
quite scarce\cite{Wu2015, Luber2015}. The first experimental ROA spectrum of a metal complex was published in 2011 by 
Johanessen \textit{et al.}\cite{Johannessen2011}, while the first theoretical ROA spectra for chiral transition metal 
complexes had already been presented in 2008 by Luber and Reiher\cite{Luber2008}. The same group then presented theoretical 
studies on chiral Cobalt complexes\cite{Luber2009, Luber2010a}, and the $\beta$ domain of rat metallothionein, which 
features one cadmium and two zinc atoms in its center\cite{Luber2010}. More recently, Humbert-Droz \textit{et al.} published 
a combined theoretical and experimental study on the ROA spectrum of $\Lambda$-tris-(ethylenediamine)-rhodium(III) 
([Rh(en)\textsubscript{3}]\textsuperscript{3+})\cite{Humbert-Droz2014}. In summary, all these studies have demonstrated 
the great potential of ROA spectroscopy for the elucidation of the structure of chiral metal complexes.

It has been found that the ROA signal can be greatly enhanced if the wavelength of the incident laser light is in resonance 
with an electronic excitation of the molecule under investigation\cite{Barron2004a}, an effect called resonance ROA (RROA).
A corresponding theory, valid in the case of resonance with a single excited electronic state, was presented in 1996 by 
Nafie\cite{Nafie1996}. This theory predicts the resulting spectra to be monosignate and the relative band intensities 
identical to the one of the corresponding Raman spectrum. These predictions were confirmed by an experimental study conducted 
by Vargek \textit{et al.}~in 1998\cite{Vargek1998}. If several excited electronic states are involved, however, the situation 
will be much more complicated; the spectra do not have to be monosignate\cite{Merten2012, Vidal2016} and de-enhancement 
effects may occur\cite{Luber2010b}. In 2007, Jensen \textit{et al.}~presented an ansatz to calculate both off- and on-resonance 
ROA spectra, which also allowed to include multiple excited electronic states\cite{Jensen2007}. Later, also Luber \textit{et 
al.}\cite{Luber2010b}~as well as Barone and coworkers\cite{Baiardi2018} presented other approaches capable of taking 
multiple excited electronic states into account.

Unfortunately, there are almost no studies investigating RROA effects in transition metal compounds (the first experimental 
RROA spectra for a chiral transition metal compound was presented by Merten \textit{et al.}~in 2010\cite{Merten2010}). 
Furthermore, detailed theoretical studies of RROA spectra involving more than one excited electronic state are largely missing.
This paper contributes to filling this gap by presenting an in-depth study of the full-resonance and near-resonance
spectra of [Rh(en)\textsubscript{3}]\textsuperscript{3+}. Thereby it also complements the detailed study published by Humbert-Droz
\textit{et al.}~on the same compound, which focused exclusively on the off-resonance case\cite{Humbert-Droz2014}.

This paper is organized as follows: First, we give a short introduction into the theoretical background of the calculation of 
(resonance) ROA spectra. Then, in Section \ref{sec:computational_methodology}, we explain the computational methodology
followed in this study. Afterwards, the results are presented in Section \ref{sec:results}. Finally, a conclusion is given 
in Section \ref{sec:conclusion}.

\section{Theoretical Background}
\label{sec:theory}

The (resonance) ROA intensity $\Delta I_p(180^{\circ})$ of a given normal mode $p$ in a backscattering geometry (in which 
case the ROA intensity is maximized\cite{Zhu2005}) can be calculated as
\begin{equation}\label{eq:roa-int}
 \Delta I_p(180^{\circ}) = I^{R}_p(180^{\circ}) - I^{L}_p(180^{\circ}) \propto \frac{1}{c} (24\beta(G')^2_p + 8\beta(A)^2_p),
\end{equation}
where $c$ is the speed of light in vacuum, $\beta(G')^2_p$ is the anisotropic invariant of the electric-dipole--magnetic-dipole 
polarizability transition tensor, and $\beta(A)^2_p$ is the anisotropic invariant of the electric-dipole--electric-quadrupole 
polarizability transition tensor (note that all these quantities are independent of the experimental setup). The latter 
two quantities are given by
\begin{equation}
 \beta(G')^2_p = \mathrm{Im}\left( \frac{\mathrm{i}}{2} (3\alpha_{\alpha\beta}
^{(p)}G_{\alpha\beta}^{\prime(p)*} - \alpha_{\alpha\alpha}^{(p)}G_{\beta\beta}^{\prime(p)*}) 
\right),
\end{equation}
and
\begin{equation}
 \beta(A)^2_p = \mathrm{Re} \left( \frac{1}{2} \omega\alpha_{\alpha\beta}^{(p)}\epsilon_{\alpha\gamma\delta}A_{\gamma\delta\beta}^{(p)*} \right),
\end{equation}
respectively. $\alpha_{\alpha\beta}^{(p)}$ is the $\alpha\beta$-component ($\alpha, \beta = \{x, y, z\}$) of the 
electric-dipole--electric-dipole polarizability for the $p$-th normal mode, while $G_{\alpha\beta}^{\prime(p)}$ is the 
$\alpha\beta$-component of the electric-dipole--magnetic-dipole polarizability for the $p$-th normal mode. The asterisk 
denotes the complex conjugate. $\omega$ is the frequency of the incident light, and $\epsilon_{\alpha\gamma\delta}$ is 
the third-rank antisymmetric unit tensor. $A_{\gamma\delta\beta}^{(p)}$ is the $\gamma\delta\beta$-component ($\gamma, 
\delta = \{x, y, z\}$) of the electric-dipole--electric-quadrupole polarizability for the $p$-th normal mode. Re and Im 
denote the real and imaginary part, respectively. Note that the Einstein summation convention has been invoked for repeated 
greek indices.

These polarizability tensors are then calculated as
\begin{equation} \alpha_{\alpha\beta}^{(p)}G_{\alpha\beta}^{\prime(p)} = \left( \frac{\partial \alpha_{\alpha\beta}}{\partial Q_p}\right)_0 \left(\frac{\partial G'_{\alpha\beta}}{\partial Q_p}\right)_0,
\end{equation}
and
\begin{equation}
 \alpha_{\alpha\beta}^{(p)}\epsilon_{\alpha\gamma\delta}A_{\gamma\delta\beta}^{(p)} =  \left( \frac{\partial \alpha_{\alpha\beta}}{\partial Q_p}\right)_0 \epsilon_{\alpha\gamma\delta} \left( \frac{\partial A_{\gamma\delta\beta}}{\partial Q_p} \right)_0
\end{equation}
where $Q_p$ is the normal mode of the $p$-th vibration and the subscript "0" indicates that the partial derivative is to 
be evaluated at the molecular equilibrium structure. 

In the short-time approximation adopted in the ansatz by Jensen \textit{et al.}\cite{Jensen2007}, $\alpha_{\alpha\beta}$, $G'_{\alpha\beta}$, and $A_{\gamma\delta\beta}$ are 
complex quantities:
\begin{equation}
 \alpha_{\alpha\beta} = \alpha_{\alpha\beta}^{(R)} + \mathrm{i}\alpha_{\alpha\beta}^{(I)},
\end{equation}
\begin{equation}
 G'_{\alpha\beta} = G_{\alpha\beta}^{\prime(R)} + \mathrm{i}G_{\alpha\beta}^{\prime(I)},
\end{equation}
and
\begin{equation}
 A_{\gamma\delta\beta} = A_{\gamma\delta\beta}^{(R)} + \mathrm{i}A_{\gamma\delta\beta}^{(I)}.
\end{equation}
The real and imaginary parts are formally calculated from a sum over the electronic states,
\begin{equation}
 \label{eq:start}
 \alpha_{\alpha\beta}^{(R)} = \sum_{n \neq 0}f^{+}(\omega_{n0}, \omega, \Gamma)\mathrm{Re}\left( \left\langle \psi_0 \left| \mu_\alpha \right| \psi_n  \right\rangle \left\langle \psi_n \left| \mu_\beta \right| \psi_0 \right\rangle \right),
\end{equation}
\begin{equation}
 \alpha_{\alpha\beta}^{(I)} = \sum_{n \neq 0}g^{-}(\omega_{n0}, \omega, \Gamma)\mathrm{Re}\left( \left\langle \psi_0 \left| \mu_\alpha \right| \psi_n  \right\rangle \left\langle \psi_n \left| \mu_\beta \right| \psi_0 \right\rangle \right),
\end{equation}
\begin{equation}
 G_{\alpha\beta}^{\prime(R)} = \sum_{n \neq 0}f^{-}(\omega_{n0}, \omega, \Gamma)\mathrm{Im}\left( \left\langle \psi_0 \left| \mu_\alpha \right| \psi_n  \right\rangle \left\langle \psi_n \left| m_\beta \right| \psi_0 \right\rangle \right),
\end{equation}
\begin{equation}
 G_{\alpha\beta}^{\prime(I)} = \sum_{n \neq 0}g^{+}(\omega_{n0}, \omega, \Gamma)\mathrm{Im}\left( \left\langle \psi_0 \left| \mu_\alpha \right| \psi_n  \right\rangle \left\langle \psi_n \left| m_\beta \right| \psi_0 \right\rangle \right),
\end{equation}
\begin{equation}
 A_{\alpha\beta\gamma}^{(R)} = \sum_{n \neq 0}f^{+}(\omega_{n0}, \omega, \Gamma)\mathrm{Re}\left( \left\langle \psi_0 \left| \mu_\alpha \right| \psi_n  \right\rangle \left\langle \psi_n \left| \Theta_{\beta\gamma} \right| \psi_0 \right\rangle \right),
\end{equation}
and
\begin{equation}
 \label{eq:end}
 A_{\alpha\beta\gamma}^{(I)} = \sum_{n \neq 0}g^{-}(\omega_{n0}, \omega, \Gamma)\mathrm{Re}\left( \left\langle \psi_0 \left| \mu_\alpha \right| \psi_n  \right\rangle \left\langle \psi_n \left| \Theta_{\beta\gamma} \right| \psi_0 \right\rangle \right).
\end{equation}
In these equations, $\psi_0$ is the electronic wave function of the ground state, while $\psi_n$ denote the electronic 
wave functions of excited states. Furthermore, we require the electric dipole operator
\begin{equation}
 \mu_\alpha = -\sum_{a}r_{a\alpha},
\end{equation}
and the magnetic dipole operator
\begin{equation}
 m_\alpha = -\frac{1}{2}\sum_al_{a\alpha}
\end{equation}
with $l_{a\alpha}$ being the $\alpha$-component of the angular momentum operator for electron $a$. The electric quadrupole 
operator
\begin{equation}
  \Theta_{\alpha\beta} = -\frac{1}{2}\sum_a(3r_{a\alpha}r_{a\beta}-\delta_{\alpha\beta}r_a^2) .
\end{equation}
The sums run over all electrons $a$. Finally, in equations (\ref{eq:start})\,--\,(\ref{eq:end}) we have the two line 
shape functions
\begin{equation}
 f^\pm(\omega_{n0}, \omega, \Gamma) = \frac{\omega_{n0} - \omega}{(\omega_{n0} - \omega)^2 + \Gamma^2} \pm \frac{\omega_{n0} + \omega}{(\omega_{n0} + \omega)^2 + \Gamma^2},
\end{equation}
and
\begin{equation}
 g^\pm(\omega_{n0}, \omega, \Gamma) = \frac{\Gamma}{(\omega_{n0} - \omega)^2 + \Gamma^2} \pm \frac{\Gamma}{(\omega_{n0} + \omega)^2 + \Gamma^2}.
\end{equation}
Here, $\omega_{n0}$ is the excitation frequency between the ground state and the $n$-th excited state, $\omega$ is the 
frequency of the incident light, and  $1/\Gamma$ can be interpreted as a lifetime of the excitations. Note that by
setting $\Gamma = 0$ in the above equations, the standard non-resonant formulation of ROA is recovered.

\section{Computational Methodology}
\label{sec:computational_methodology}

\begin{figure}[H]
\begin{center}
\includegraphics[scale=0.3]{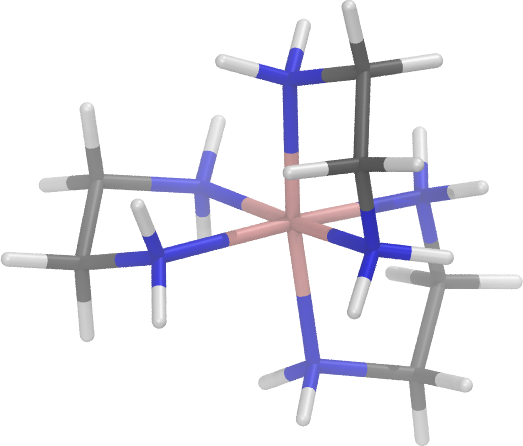}
\end{center}
\caption{\label{fig:struct}\small Structure of [Rh(en)\textsubscript{3}]\textsuperscript{3+} investigated in this work.}
\end{figure}

In their study\cite{Humbert-Droz2014}, Humbert-Droz \textit{et al.}~analyzed in detail the different configurations and 
conformations possible for [Rh(en)\textsubscript{3}]\textsuperscript{3+}. Therefore, we will not attempt to repeat such
an analysis in this work. Comparison with experimental spectra showed that the so-called $\Lambda(\delta\delta\delta)$ 
conformation (see Fig.~\ref{fig:struct}; \textit{cf.}~Ref.~\citenum{connelly2005} for more information about this nomenclature) 
to be the dominant conformation\cite{Humbert-Droz2014}. Therefore, we will use this sole conformation throughout our study 
and refer to it without the prefix ``$\Lambda(\delta\delta\delta)$''.

The structure of [Rh(en)\textsubscript{3}]\textsuperscript{3+} was first fully optimized in C\textsubscript{1} symmetry 
(see the Supporting Information for the Cartesian coordinates of the optimized structure). This structure optimization 
and all subsequent calculations were done with a development version of \textsc{NWChem} (revision 29384)\cite{Valiev2010}, 
using density functional theory in its Kohn--Sham formulation\cite{Kohn1965}.  (It has been verified that the ROA spectra 
obtained with this specific version of \textsc{NWChem} are identical to the ones obtained from the officially released 
\textsc{NWChem} 6.8.1 by recalculating the spectra at 800\,nm and 307.66\,nm incident wavelength, respectively.) In all 
calculations, the PBE0 exchange--correlation functional\cite{Adamo1999} and the def2-TZVP basis set\cite{Weigend2005} 
(for all atoms) were chosen. For Rhodium, the ECP-28 effective core potential\cite{Andrae1990} was employed in order to 
take scalar relativistic effects into account. Furthermore, advantage was taken of the resolution-of-the-identity technique 
for the Coulomb integrals using the corresponding auxiliary basis sets\cite{Weigend2006}. Finally, the D3 semiempirical 
dispersion corrections\cite{Grimme2010} were adopted (with the zero-damping function).

All ROA spectra were calculated in the double harmonic approximation with the program suite \textsc{MoViPac}\cite{Weymuth2012}, 
relying on a seminumerical differentiation scheme with a three-step central difference formula (adopting a step size of 
0.01\,bohr). The necessary energy gradients and polarizability tensors were calculated analytically with the version of 
\textsc{NWChem} mentioned above relying on the ansatz by Jensen \textit{et al.}\cite{Jensen2007}. For all spectra, a 
value of 0.008\,a.u.~was adopted for the damping parameter $\Gamma$ 
(we refer to the Supporting Information for a comparison of different damping values). Furthermore, great care was taken 
to ensure origin independence of the spectra by placing all structures close to the origin and using gauge-including 
atomic orbitals; note that for resonance ROA calculations, gauge-including atomic orbitals alone cannot always guarantee 
complete origin independence\cite{Vidal2016} (see the Supporting Information for calculations verifying the gauge independence 
in our case). Electronic (singlet) excitations were calculated with time-dependent density functional theory employing the 
methodology detailed above. This yielded the lowest excitation at 307.66\,nm. Therefore, this wavelength was used in order 
to calculate the full-resonance spectrum. In order to investigate how this spectrum changes with different wavelengths, 
spectra were also calculated for the following wavelengths: 290\,nm, 303\,nm, 305\,nm, 310\,nm, 315\,nm, 320\,nm, 325\,nm, 
330\,nm, 350\,nm, 370\,nm, 410\,nm, 500\,nm, 600\,nm, 700\,nm, and 800\,nm. Solvent effects were not considered in
this study, as this allows us to analyze the genuine ROA spectra and of [Rh(en)\textsubscript{3}]\textsuperscript{3+}
itself and the resonance effects stemming from this isolated structure. Furthermore, as we will see below, solvent effects 
do not appear to play a major role in this ROA spectrum. A complete input file used to calculate the ROA spectra is given 
in the Supporting Information.

All line spectra were convoluted with a Lorentzian band having a full width at half maximum of 15\,cm$^{-1}$. This line
broadening as well as the subsequent analysis of all ROA spectra was carried out with \textsc{Mathematica} 11.2.0\cite{mathematica11_2}.
Plots of ROA spectra were created with \textsc{Matplotlib} 2.2.3\cite{Hunter2007}. In these plots, the intensities of the 
line spectra were always scaled by a factor of 0.04. Pictures of molecular structures and molecular orbitals were generated 
with \textsc{Vmd} 1.9.2\cite{Humphrey1996, tachyon}. Representations of normal modes were produced with \textsc{Jmol} 
14.6.4\cite{jmol}.

\section{Results}
\label{sec:results}

\subsection{Electronic Excitations}
\label{subsec:excitations}

To assess at which wavelength a full-resonance ROA spectrum can be calculated, we need to know the energies of the 
electronic excitations. The calculated energies and wavelengths of the lowest five electronic (singlet) excitations are
collected in Table~\ref{tab:energies}. 

\begin{table}[H]
\renewcommand{\baselinestretch}{1.0}
\renewcommand{\arraystretch}{1.0}
\caption{\label{tab:energies}\small Excitation energies $E$ and corresponding wavelengths $\lambda$ of the lowest five 
electronic excitations as well as the molecular orbital transitions involved with a contribution larger than 0.4\,a.u.~to 
the excitation vector (transitions with a contribution larger than 0.6\,a.u.~are printed in boldface).} 
\begin{center}
\begin{tabular}{l r r l} \hline \hline
No.   & $E$ / eV   & $\lambda$ / nm   & Dominating Transitions                                                \\
\hline 
1     & 4.0299     & 307.6607         & \textbf{HOMO $\rightarrow$ LUMO},     \\
      &            &                  & HOMO\,$-$\,1 $\rightarrow$ LUMO\,+\,1, \\
      &            &                  & HOMO\,$-$\,2 $\rightarrow$ LUMO \\
2     & 4.0301     & 307.6455         & \textbf{HOMO $\rightarrow$ LUMO\,+\,1}, \\
      &            &                  & HOMO\,$-$\,1 $\rightarrow$ LUMO, \\
      &            &                  & HOMO\,$-$\,2 $\rightarrow$ LUMO\,+\,1 \\
3     & 4.0864     & 303.4069         & \textbf{HOMO\,$-$\,1 $\rightarrow$ LUMO\,+\,1},\\
      &            &                  & \textbf{HOMO\,$-$\,2 $\rightarrow$ LUMO}             \\
4     & 4.6548     & 266.3577         & n/a                                                                    \\
5     & 4.6551     & 266.3406         & n/a                                                                    \\
\hline
\hline
\end{tabular}
\renewcommand{\baselinestretch}{1.0}
\renewcommand{\arraystretch}{1.0}
\end{center}
\end{table}

The lowest excitation occurs at a wavelength of 307.66\,nm. We therefore adopted this wavelength in the calculation of 
the full-resonance ROA spectrum. However, we can also see from Table~\ref{tab:energies} that the second excitation occurs 
at almost the same energy. Hence, both states will simultaneously be involved in resonance effects. As we will understand 
below, this fact leads to very interesting effects in the ROA spectra. Moreover, as the third lowest state (at 303.41\,nm 
above the ground state) is also very close to the aforementioned two states, it is likely that also this state will usually 
be involved in resonance effects, albeit to a somewhat lesser extent. Resonance effects with states beyond the third lowest 
excited state will be negligible, since the excitation wavelengths are at least 50\,nm larger compared to the lowest 
excitation (this is also indicated by the findings of Ref.~\citenum{Luber2010b}).

\begin{figure}[H]
\begin{center}
\includegraphics[scale=0.18]{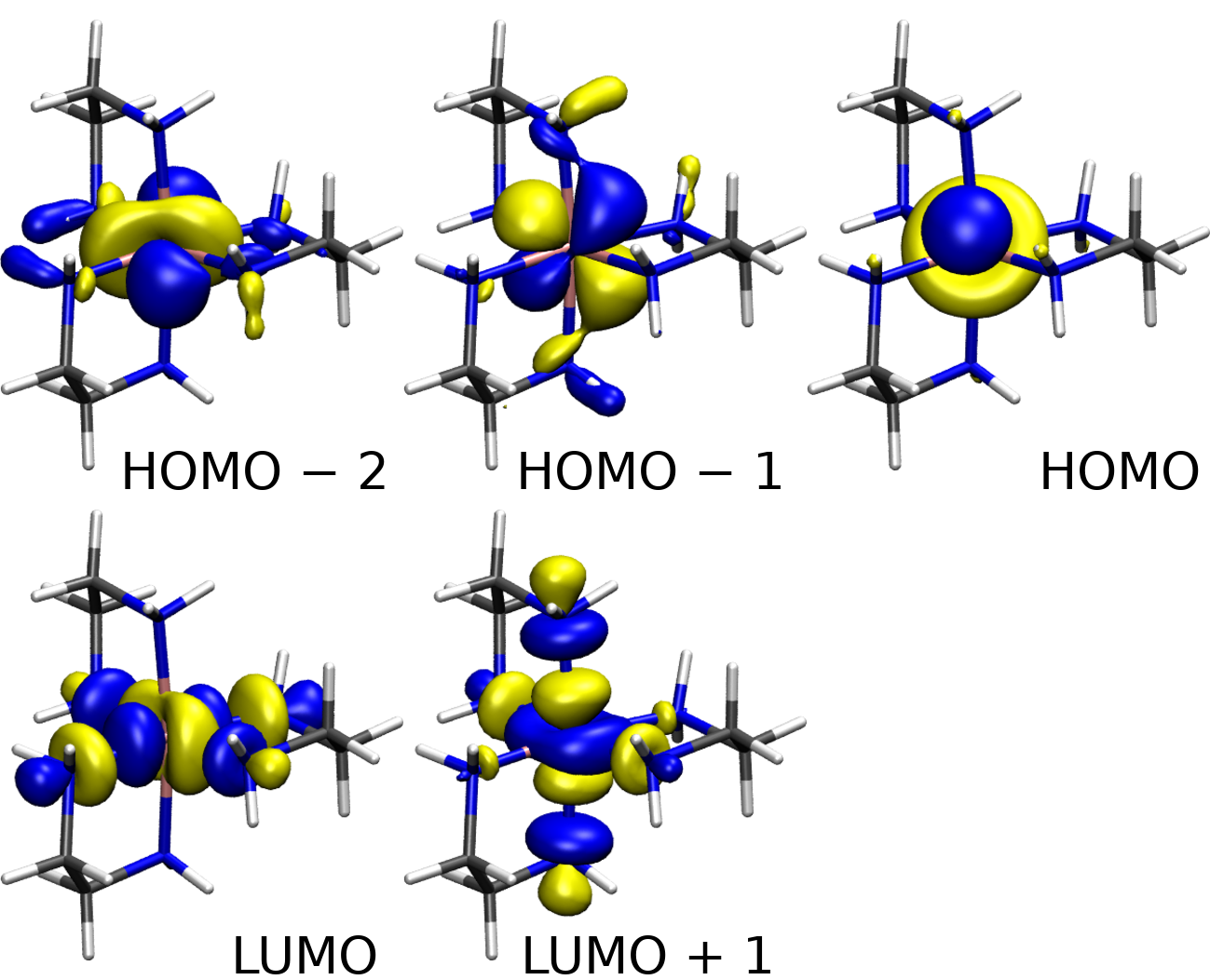}
\end{center}
\caption{\label{fig:orbitals}\small Molecular orbitals of [Rh(en)\textsubscript{3}]\textsuperscript{3+} involved in the
lowest three electronic excitations.}
\end{figure}

It is also instructive to analyze the molecular orbitals (MOs) involved in the three lowest electronic excitations. The 
dominant MO transitions involved in these excitations are also listed in Table~\ref{tab:energies} and the respective MOs
are shown in Fig.~\ref{fig:orbitals}. The first excitation is mostly a transition from the highest occupied MO (HOMO) to 
the lowest unoccupied MO (LUMO). While the HOMO is localized at the metal center, the LUMO is more extended, covering also 
four of the nitrogen atoms. One can therefore interpret this transition as moving some electronic charge from the rhodium 
atom to the nitrogen atoms. This interpretation remains valid when considering also the other two dominating transitions, 
namely the transitions between HOMO\,$-$\,1 and LUMO\,+\,1, and between HOMO\,$-$\,2 and LUMO. HOMO\,$-$\,2 and HOMO\,$-$\,1 
are mostly localized at the central rhodium atom, while LUMO\,+\,1 is, like the LUMO, more extended, also covering several 
nitrogen atoms. Following the same argumentation, we find that also the second and third lowest transition represent a 
movement of charge from the metal center to the adjacent nitrogen atoms.

\subsection{ROA Spectra}

After having analyzed the electronic excitations playing a role in the (resonance) ROA spectra, we are now in a position 
to investigate these spectra.

\begin{figure}[H]
\begin{center}
\includegraphics[scale=0.4]{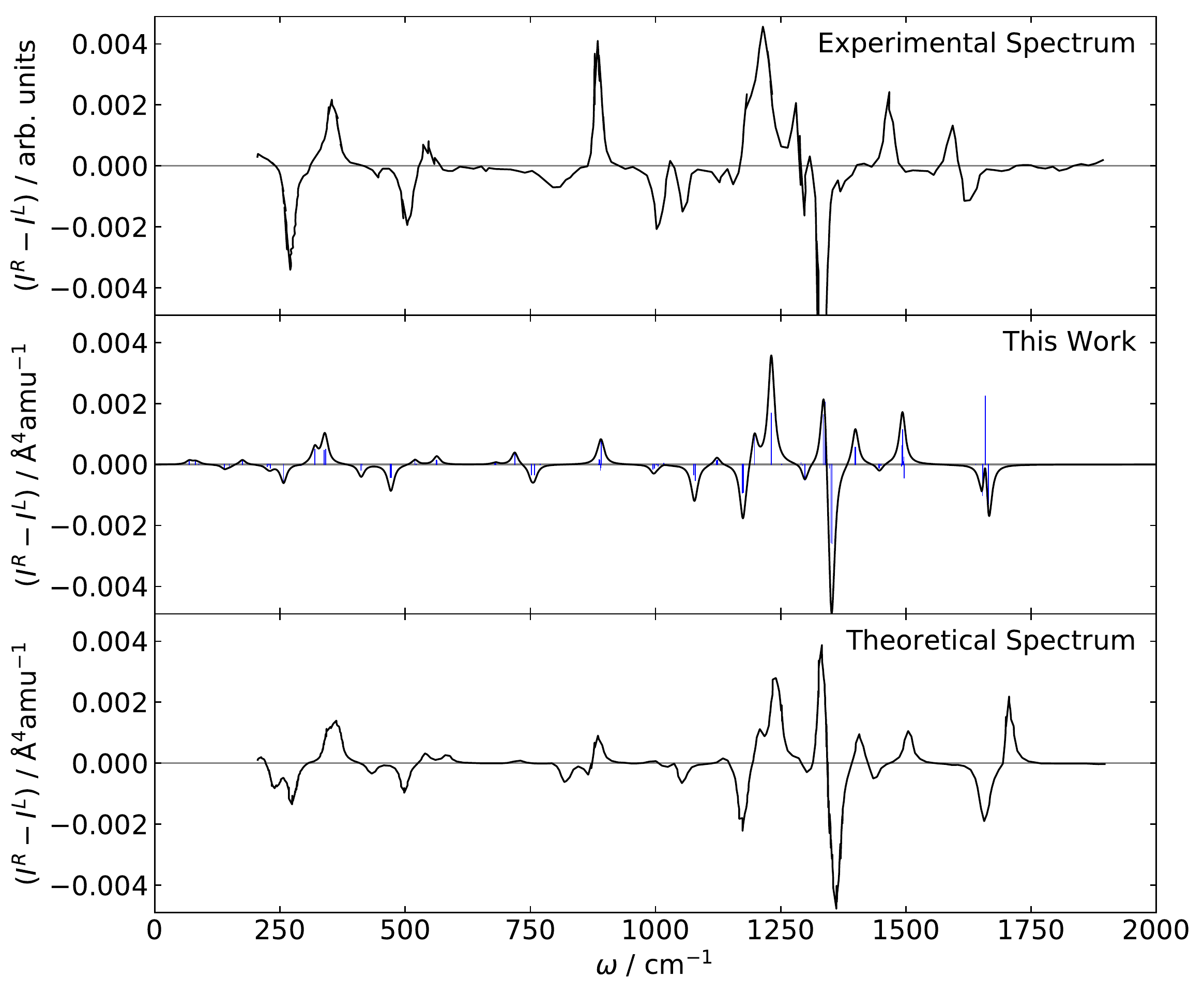}
\end{center}
\caption{\label{fig:comparison}\small Comparison of the experimental (top panel) and theoretical (bottom panel) off-resonance 
spectra of [Rh(en)\textsubscript{3}]\textsuperscript{3+} as obtained by Humbert-Droz \textit{et al.}\cite{Humbert-Droz2014}
to the off-resonance spectrum obtained in this work (middle panel).}
\end{figure}

First, we compare our results to the ones obtained by Humbert-Droz \textit{et al.}\cite{Humbert-Droz2014}. In 
Fig.~\ref{fig:comparison}, we show both the experimental as well as the theoretical off-resonance spectrum obtained by 
these authors and compare it to the off-resonance spectrum obtained in this work using an excitation wavelength of 800\,nm. 
In the work by Humbert-Droz \textit{et al.}\cite{Humbert-Droz2014}, an excitation wavelength of 532\,nm was adopted. As 
we have elaborated in the previous section, this wavelength is in fact far away from any electronic excitations. We will 
see below that for wavelengths longer than 350\,nm, resonance effects do not play a major role anymore in the ROA spectrum
of [Rh(en)\textsubscript{3}]\textsuperscript{3+}. The experimental spectrum is, therefore, effectively an off-resonance
spectrum and it is possible to compare the spectra shown in Fig.~\ref{fig:comparison} to each other, even though they have 
been obtained with different excitation wavelengths. The apparent ruggedness of the two spectra taken from Ref.~\citenum{Humbert-Droz2014} is 
in part due to the fact that it was extracted directly from the paper by Humbert-Droz \textit{et al.}\cite{Humbert-Droz2014},
which introduced minor artifacts. Also note that Humbert-Droz \textit{et al.}~adopted a different computational methodology;
most notably, instead of the PBE0 density functional, they used the B3LYP\cite{Becke1993, Lee1988, Vosko1980} exchange--correlation 
functional (as implemented in \textsc{Gaussian}\cite{gaussian09}). Therefore, 
some differences between the theoretical spectra are to be expected.

As we can see, the three spectra agree well with each other, in particular the two theoretical spectra. Below
1000\,cm$^{-1}$, the only notable differences between these two spectra are slightly different band shapes in the region
between 200\,cm$^{-1}$ and 300\,cm$^{-1}$, and the +/$-$ couplet at about 750\,cm$^{-1}$, which is visible in the spectrum 
calculated by us, but not in the theoretical spectrum obtained by Humbert-Droz \textit{et al.}\cite{Humbert-Droz2014}.
Instead of such a couplet, however, we find two weakly negative peaks at about 800\,cm$^{-1}$ which cannot be identified
in our spectrum. Also above 1000\,cm$^{-1}$, the two theoretical spectra agree well. The only difference is in the region 
between 1650\,cm$^{-1}$ and 1750\,cm$^{-1}$. Here, we find two negative bands in our spectrum, while there is a rather 
strong $-$/+ couplet in the spectrum reported in Ref.~\citenum{Humbert-Droz2014} (note that in the experimental spectrum, 
there is a +/$-$ couplet in this spectral range). However, also in the spectrum obtained in our study, we find a normal 
mode with a strongly positive ROA intensity in this wavenumber range. However, it is completely covered up by the near-lying 
modes which are all associated with negative ROA intensity. If this mode would be shifted to slightly larger wavenumbers, 
also our spectrum would feature a $-$/+ couplet. This highlights the fact that the overall band shape obtained in ROA 
spectra is highly sensitive to such comparatively small changes (\textit{cf.}, also Ref.~\citenum{Herrmann2006}). Therefore, 
in all ROA spectra reported in this study, we will always also show the individual normal modes.

At wavenumbers below 1000\,cm$^{-1}$, the experimentally measured intensities are clearly larger than the calculated ones, 
but the wavenumbers of the measured peaks matches those of the calculated peaks almost perfectly. The +/$-$ couplet at 
about 750\,cm$^{-1}$, which we find in our spectrum, cannot be identified in the experimental spectrum. Instead, the 
experimental spectrum features a broad, negative band at around 800\,cm$^{-1}$. This is consistent with what we find in 
the theoretical spectrum obtained by Humbert-Droz \textit{et al.}\cite{Humbert-Droz2014}. At wavenumbers above 
1000\,cm$^{-1}$, the relative intensity ratio between the two strongly positive peaks at about 1200\,cm$^{-1}$ and 
1300\,cm$^{-1}$, respectively, is correct reproduced in our spectrum, while it is reversed in the spectrum given in
Ref.~\citenum{Humbert-Droz2014}. However, these are minor differences, and the overall agreement of all three spectra is 
very satisfactory.

\begin{figure}[H]
\begin{center}
\includegraphics[scale=0.4]{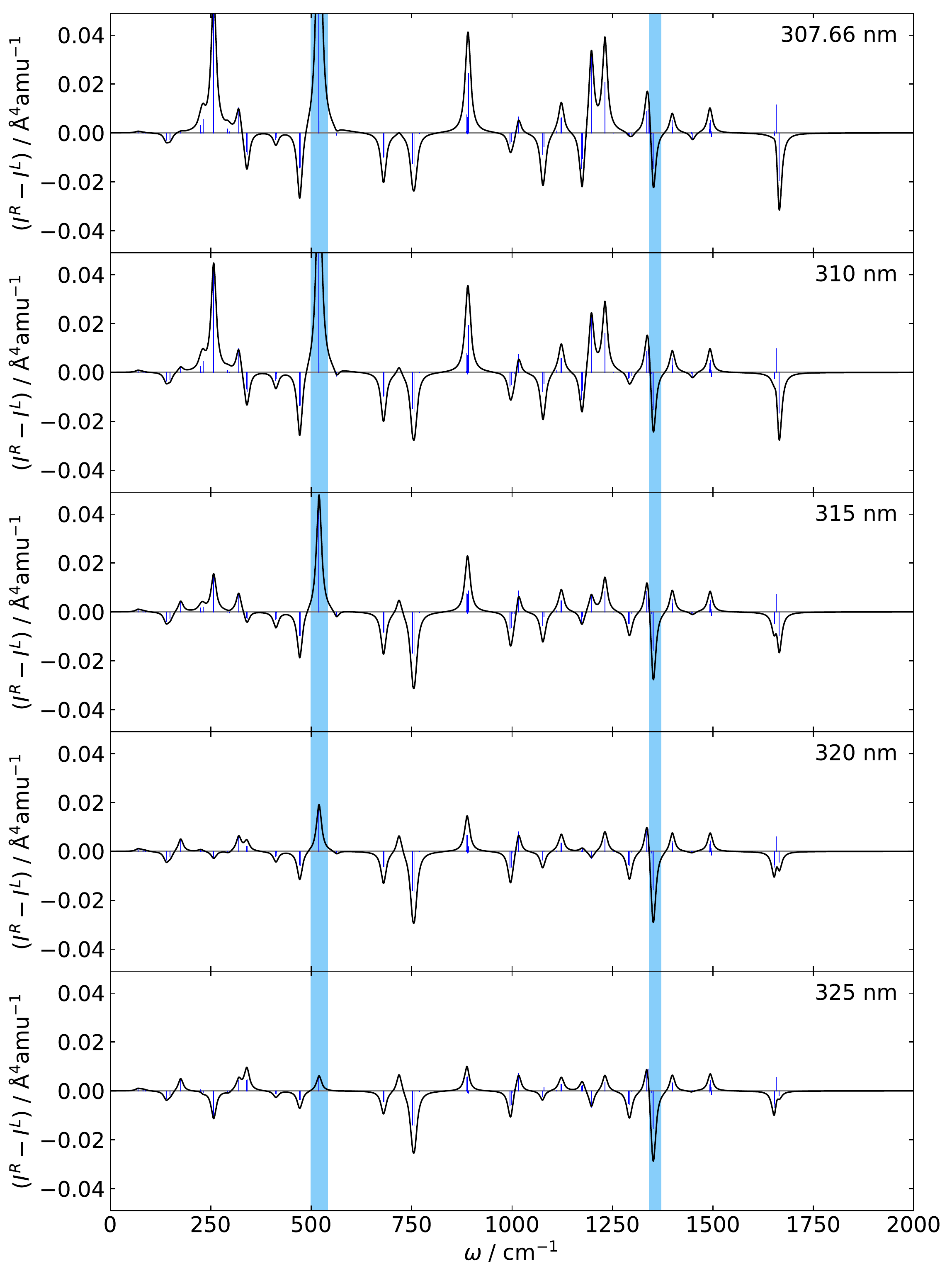}
\end{center}
\caption{\label{fig:spectra_near_res}\small ROA spectra for five different wavelengths at or close to resonance with
the first three excited electronic states (in descending order from top to bottom). The two bands primarily analyzed in
this study are highlighted.}
\end{figure}

\begin{figure}[H]
\begin{center}
\includegraphics[scale=0.4]{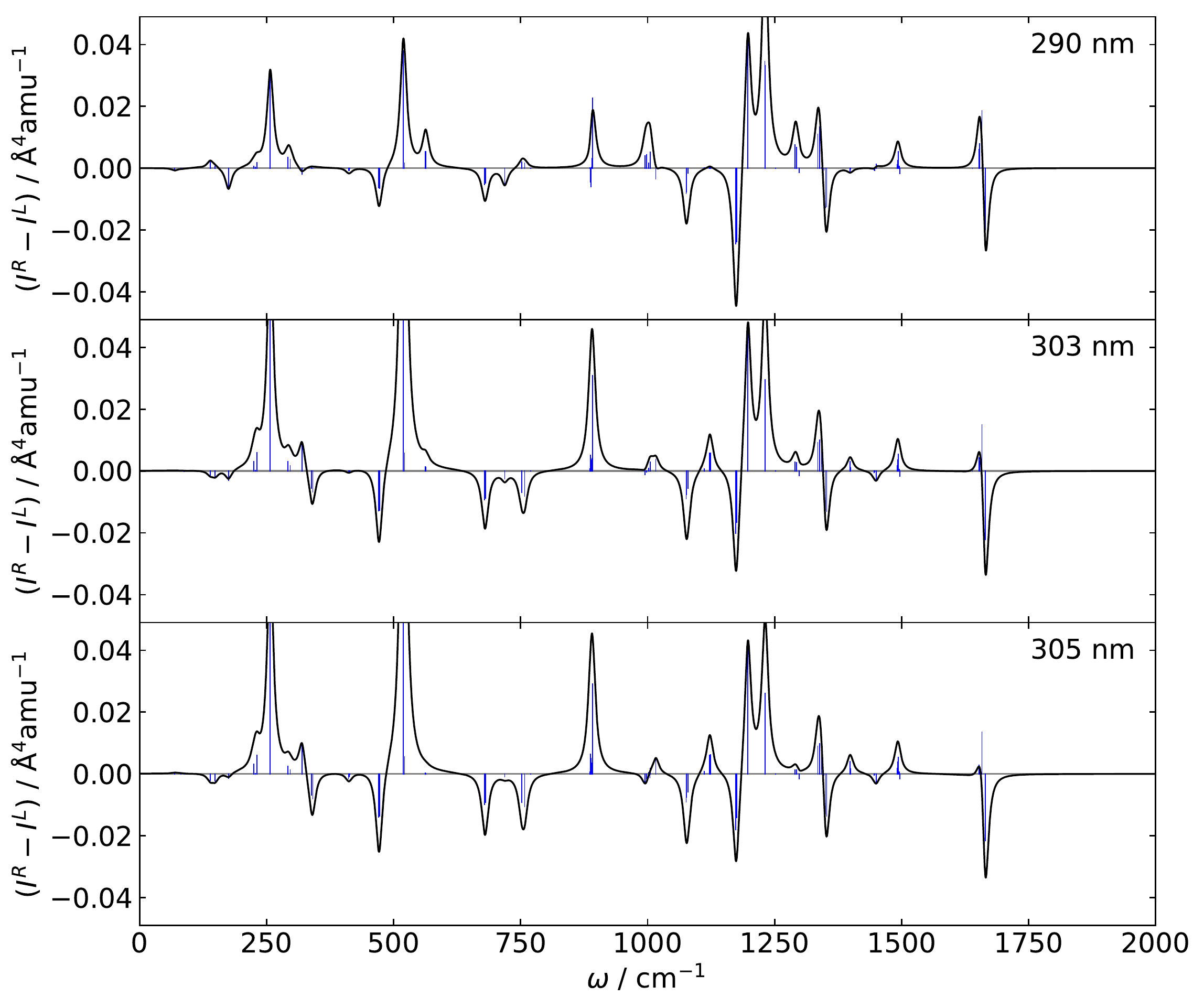}
\end{center}
\caption{\label{fig:spectra_pre_res}\small ROA spectra for three pre-resonance wavelengths.}
\end{figure}

Now, we can analyze all our ROA spectra and how they are affected by resonances with electronic states. The individual
spectra are shown in Figs.~\ref{fig:spectra_near_res}\,--\,\ref{fig:spectra_pre_res}. The full-resonance spectrum (at 
307.66\,nm, see top spectrum in Fig.~\ref{fig:spectra_near_res}) shows bisignate peaks, \textit{i.e.}, it is not monosignate, 
as is often the case for RROA spectra. This is because RROA spectra are only monosignate in the case of resonance with 
exactly one excited electronic state. In our case, however, we have seen above that the three lowest electronic excitations 
are so close to each other in energy that they all contribute to the resonance effects observed. It is well known that in 
such situations, bisignate RROA spectra can result\cite{Merten2012, Vidal2016, Luber2010b}. The intensity of the 
full-resonance spectrum reaches absolute values of up to 0.12\,\AA{}$^4$\,amu$^{-1}$ (at 520\,cm$^{-1}$). This is larger 
than the maximum intensity in the case of the spectrum recorded at 800\,nm (which is a pure off-resonance spectrum) where 
the maximum absolute intensity reaches only about 0.005\,\AA{}$^4$\,amu$^{-1}$. We  therefore clearly see an enhancement 
of the ROA intensity through the resonance with electronically excited states, as is expected.

Increasing the excitation wavelength from 307.66\,nm to 310\,nm yields a spectrum which looks very similar to the 
full-resonance spectrum (see the second spectrum from above in Fig.~\ref{fig:spectra_near_res}). All band shapes and 
relative intensity differences are basically the same, but the overall intensity is reduced, reaching a maximum value of 
about 0.095\,\AA{}$^4$\,amu$^{-1}$ at 520\,cm$^{-1}$. Increasing the excitation wavelength to 315\,nm, we find that the 
intensity is further reduced; the maximum value is now roughly 0.048\,\AA{}$^4$\,amu$^{-1}$ (still at 520\,cm$^{-1}$). 
Relative intensity differences are now strong enough to have a visible influence on the overall band shapes. For example, 
the band at 1660\,cm$^{-1}$ does no longer appear as a single sharp peak, but is split into two peaks.

In the ROA spectrum at 320\,nm, the maximum intensity is no longer found for the band at 520\,cm$^{-1}$ but for the one at 
about 760\,cm$^{-1}$. Its absolute intensity (about 0.03\,\AA{}$^4$\,amu$^{-1}$) is roughly as large as the one for the 
peak at about  1350\,cm$^{-1}$. When compared to the full-resonance spectrum at 307.66\,nm, the overall spectrum has 
changed significantly. While certain spectral regions (\textit{e.g.}, between 1400\,cm$^{-1}$ and 1500\,cm$^{-1}$) are 
almost the same in both spectra, other regions (\textit{e.g.}, between 200\,cm$^{-1}$ and 300\,cm$^{-1}$) differ 
significantly. In the case of the full-resonance spectrum, we see a strongly positive band between 200\,cm$^{-1}$ and 
300\,cm$^{-1}$. The spectrum at 320\,nm features a very weak positive peak and a slightly stronger, negative band in the 
same spectral region. Also, at about 700\,cm$^{-1}$, we can identify a positive band in the spectrum at 320\,nm. This 
was completely masked in the full-resonance spectrum (since the corresponding normal mode has a much weaker, albeit 
also positive, intensity in the full-resonance case). Furthermore, also between 1200\,cm$^{-1}$ and 1300\,cm$^{-1}$ the 
two spectra differ strongly. In the full-resonance case, we find here two strongly positive peaks, followed by a weaker 
negative one. At 320\,nm, we find only one positive peak, much less intense, while the negative peak has gained intensity. 
The normal mode at 1252\,cm$^{-1}$, causing the second strongly positive peak in the spectrum at 307.66\,nm, has almost
zero ROA intensity at 320\,nm, which is why it can no longer be clearly identified in the corresponding spectrum.
The spectrum with an excitation wavelength of 325\,nm looks very similar to the one at 320\,nm, but we can still find a 
few distinct changes. For example, the most intense peak is now found at about 1350\,cm$^{-1}$. The band at about 
520\,cm$^{-1}$ has further lost intensity.

\begin{figure}[H]
\begin{center}
\includegraphics[scale=0.38]{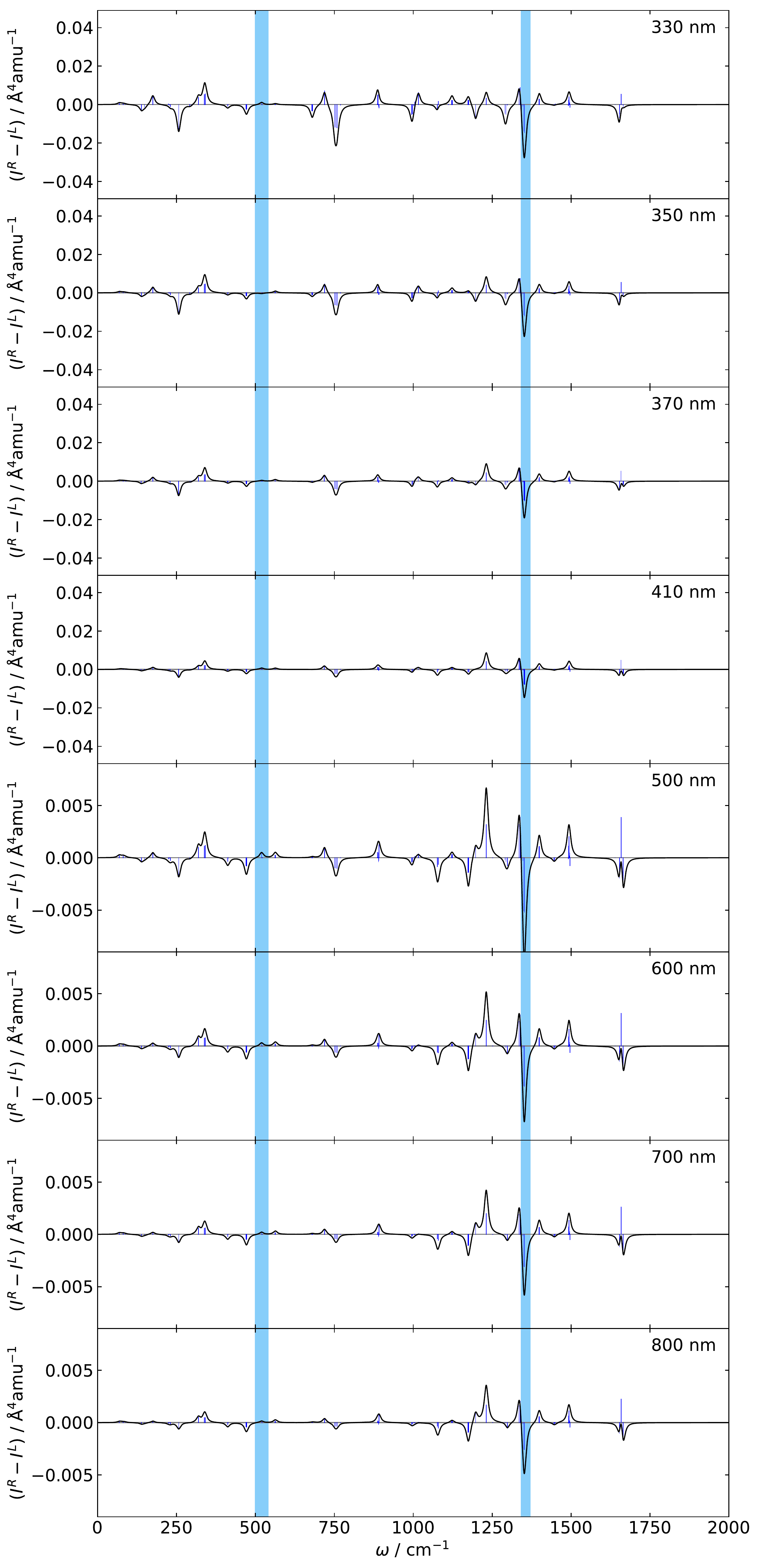}
\end{center}
\caption{\label{fig:spectra_off_res}\small ROA spectra for different off-resonant wavelengths. Note that for wavelengths
larger than 410\,nm, a smaller intensity range is plotted for the sake of clarity. The two bands primarily analyzed in
this study are highlighted.}
\end{figure}

At an excitation wavelength of 330\,nm, the overall intensity is further reduced (\textit{cf.}, Fig.~\ref{fig:spectra_off_res}). 
We can still find some significant changes in the overall band shape. In particular, the positive band at 520\,cm$^{-1}$ 
has almost completely vanished at 330\,nm. At an excitation wavelength of 350\,nm, one can see that the overall intensity 
continues to be reduced. Compared to the spectrum at 330\,nm, however, the overall band shape is conserved. In all 
spectra with excitation wavelengths larger than 350\,nm, we find well-conserved band shapes. Therefore, one can say that 
already at 350\,nm, the shape of the RROA spectrum is identical to the one of a  pure off-resonance spectrum at 800\,nm. 
However, the overall intensity is continuously decreasing with increasing excitation wavelength. This might be due to 
very weak resonance effects (absorption tail).

An interesting observation can be made for the normal mode at around 720\,cm$^{-1}$. We find that the intensity of this 
mode increases when moving away from resonance, until the highest intensity is observed at 320\,nm. For larger 
wavelengths, the intensity decreases again. When examining the intensities of the individual normal modes closer, 
we find that such a behavior is found for many of the normal modes (see Supporting Information). The reason for this 
intricate behavior is not fully clear. However, it is known that resonance with multiple excited states can lead to 
de-enhancement effects and concomitantly complex line shapes\cite{Luber2010b}. We could hypothesize that the third 
excited electronic state has a de-enhancing effect on the ROA spectrum. At longer wavelengths, the third excited state 
(303\,nm excitation wavelength) is less important than the first two, and concomitantly its de-enhancing effect should 
be diminished. It is, therefore, interesting to see  how the spectrum looks like at shorter wavelengths, when this third 
state should be more important.

Hence, we also calculated spectra at 305\,nm, 303\,nm, and 290\,nm (see Fig.~\ref{fig:spectra_pre_res}). Overall, all 
these spectra have a rather large resemblance to the spectrum at full resonance, but it can be argued (based on visual 
inspection), that the spectrum at 325\,nm is more similar to the spectrum at 307.66\,nm than the spectrum at 290\,nm (note 
that both off-resonant wavelengths have roughly the same distance to the resonant wavelength). Therefore, we might conclude 
that, as expected, the third excited state contributes much more to the spectrum at 290\,nm and therefore leads to bigger 
changes. Even though the spectrum at 305\,nm looks very similar to the spectrum at full resonance, some distinct differences 
can be identified. Most prominently, we can identify a weakly positive band at around 1650\,cm$^{-1}$, which is not visible 
at full resonance (although also in this case, there are normal modes in this spectral region associated with positive ROA 
intensity). This positive band is further accentuated at smaller wavelengths, and at 290\,nm we find a strong $+/-$ couplet 
in this region. When comparing the spectrum at 290\,nm to the one at 307.66\,nm, we find many differences. While many bands 
have decreased in intensity (as one would expect based on our hypothesis of a de-enhancing effect of the third lowest 
electronic excitation), there are also many peaks which grow in intensity. For example, the negative peak at around 
1180\,cm$^{-1}$ and the two positive peaks between 1200\,cm$^{-1}$ and 1250\,cm$^{-1}$. Hence, the third excited state 
has not always de-enhancing effects on all normal modes, but the relation appears to be more complicated.

\begin{figure}[H]
\begin{center}
\includegraphics[scale=0.25]{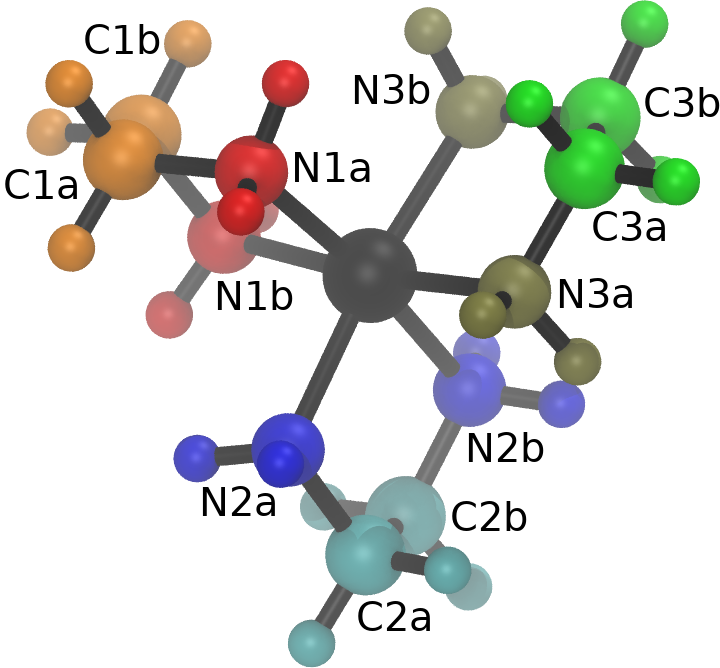}
\end{center}
\caption{\label{fig:gcm_partitioning}\small Groups of [Rh(en)\textsubscript{3}]\textsuperscript{3+} adopted for all 
group coupling matrices.}
\end{figure}

We have seen that the band at 520\,cm$^{-1}$ changes significantly when increasing the excitation wavelength; starting 
as the strongest peak in the entire spectrum, it becomes continuously weaker until it vanishes at 350\,nm. Moreover, 
the negative peak at about 1350\,cm$^{-1}$ is more or less unaffected. It is interesting, therefore, to investigate these 
two peaks more closely. An analysis in terms of so-called group coupling matrices (GCMs) as proposed by Hug\cite{Hug2001}
provides an appealing means of assigning the total ROA intensity of a given normal mode to parts of a molecule. For such 
an analysis, the proper choice of these parts is crucial. In our case, we adopt the rather fine-grained partitioning of 
[Rh(en)\textsubscript{3}]\textsuperscript{3+} as shown in Fig.~\ref{fig:gcm_partitioning}; note that also Humbert-Droz 
\textit{et al.}~have found this partitioning to be useful for some of the normal modes they analyzed\cite{Humbert-Droz2014}.

\begin{figure}[H]
\begin{center}
\includegraphics[scale=0.2]{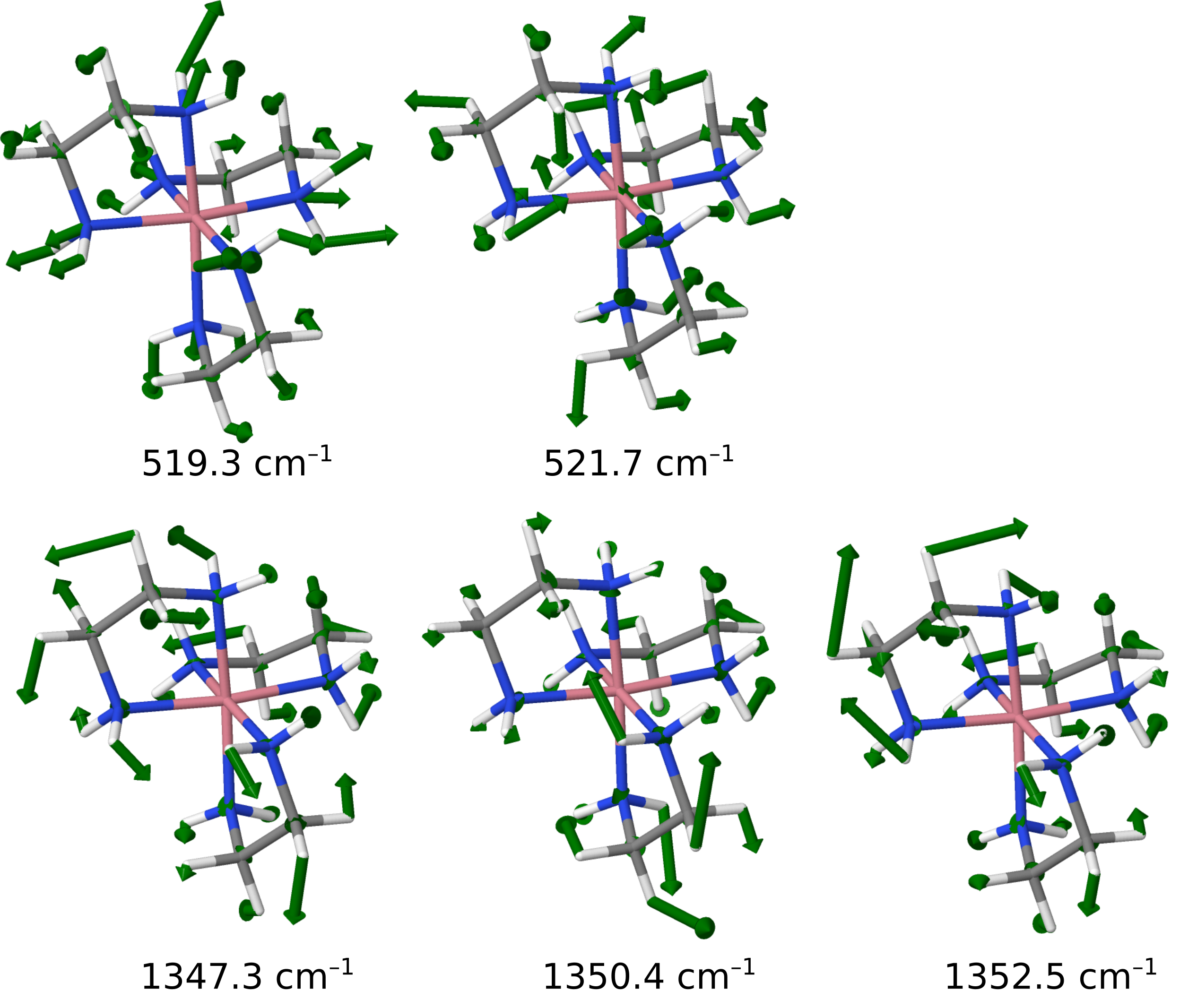}
\end{center}
\caption{\label{fig:normal_modes}\small Normal modes selected for closer analysis. The wavenumber of each normal mode is
given below its representation.}
\end{figure}

\begin{figure}[H]
\begin{center}
\includegraphics[scale=0.18]{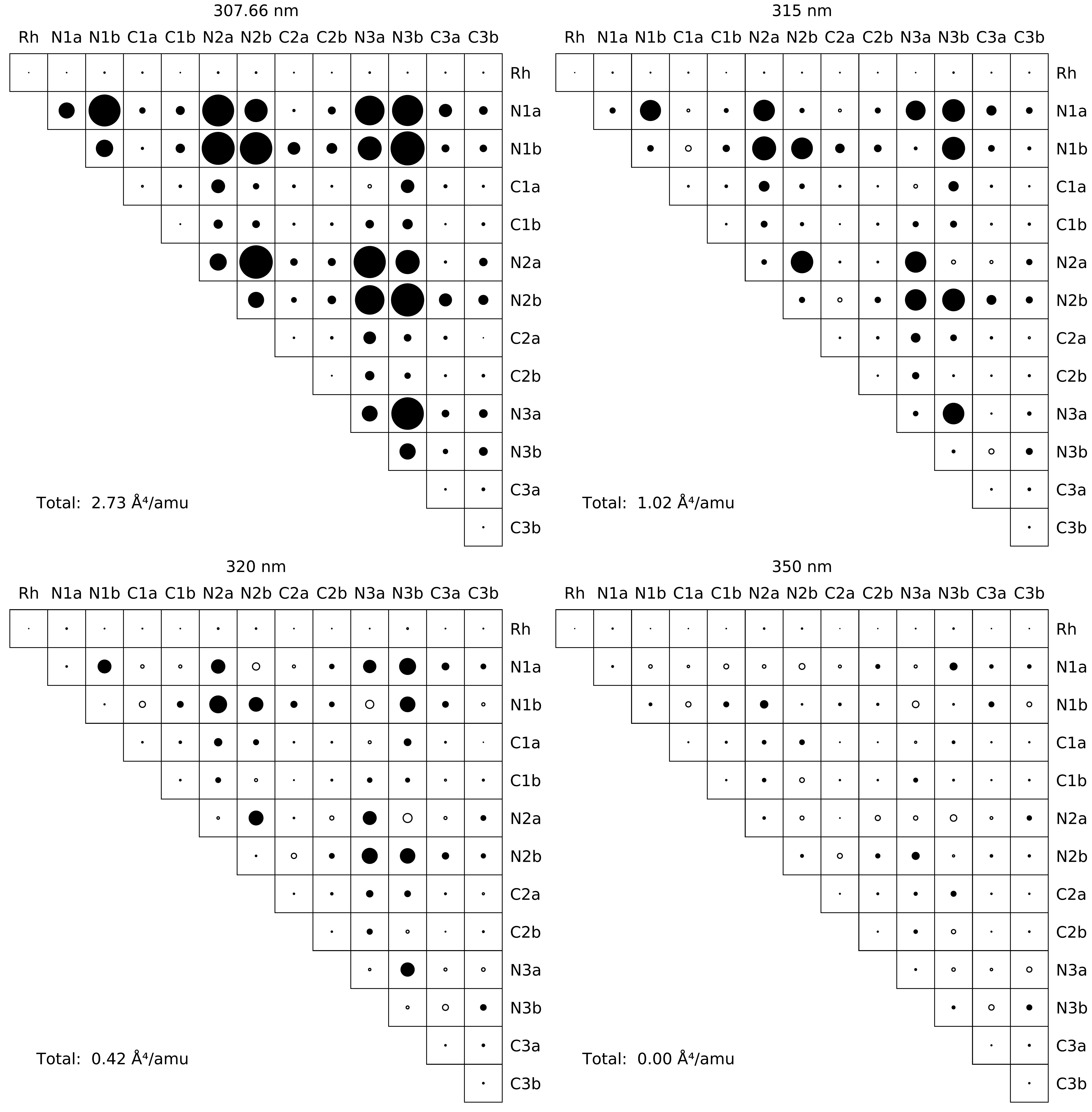}
\end{center}
\caption{\label{fig:gcm_519}\small Group coupling matrices for the normal mode at 519.3\,cm$^{-1}$ for four different 
excitation wavelengths (top left: 307.66\,nm; top right: 315\,nm; bottom left: 320\,nm; bottom right: 350\,nm). Positive 
contributions are represented by filled circles, while negative contributions are drawn as empty circles. The radius of 
the circle is proportional to the size of the corresponding GCM matrix element.}
\end{figure}

The band at about 520\,cm$^{-1}$ is composed of two normal modes.  The first normal mode (at 519.3\,cm$^{-1}$) represents 
a symmetric stretching vibration of all six Rh--N bonds, while the second mode (at 521.7\,cm$^{-1}$) describes a more 
complicated torsional motion of the three ligands (see Fig.~\ref{fig:normal_modes}). To a good approximation, we can 
neglect the contribution of this second mode to the positive band at around 520\,cm$^{-1}$, since its intensity is always 
about an order of magnitude smaller than the one of the first mode. We are therefore left with the analysis of the mode
at 519.3\,cm$^{-1}$. Its GCMs for different excitation wavelengths are shown in Fig.~\ref{fig:gcm_519}.

As we can see, at 307.66\,nm almost all parts of the molecule contribute to the positive ROA intensity. Only the coupling 
between the CH\textsubscript{2} group C1a and NH\textsubscript{2} group N3a leads to weakly negative ROA intensity, but 
this can safely be neglected. The biggest contributions to the overall ROA intensity are clearly given by the six 
NH\textsubscript{2} groups as well as couplings between them. In fact, these couplings lead to stronger contributions 
than the individual NH\textsubscript{2} groups provide, and couplings between adjacent NH\textsubscript{2} groups are not 
stronger than couplings between distant NH\textsubscript{2} groups, but not all couplings are equally strong: The coupling 
of N1a and N2b, N1b and N3a, and N2a and N3b are clearly weaker than the other couplings. From Fig.~\ref{fig:gcm_partitioning},
we see that these combinations of NH\textsubscript{2} groups are always forming an N--Rh--N angle of about 180$^\circ$,
while for any other two NH\textsubscript{2} groups, the N--Rh--N angle is about 90$^\circ$. Hence, we conclude that
the weaker couplings must be due to this specific spatial arrangement.

For the larger excitation wavelength of 315\,nm, these strong contributions have become significantly weaker, but they 
still play the dominant role. In particular, the "weak" couplings between NH\textsubscript{2} groups (\textit{i.e.}, the 
couplings between N1a and N2b, between N1b and N3a, and between N2b and N3a) are greatly reduced, being now not more important 
than any of the other couplings. Several couplings leading to negative ROA intensity are visible now, most notably the 
coupling between N2a and N3b (which contributed much positive ROA intensity at full resonance).
 
At 320\,nm, the couplings between the NH\textsubscript{2} groups are further reduced, and the intensity which the individual 
NH\textsubscript{2} groups provide directly (\textit{i.e.}, not through couplings) can be safely neglected. All weak 
NH\textsubscript{2} couplings are now clearly negative; in fact, they are the most negative contributions of the entire 
GCM. Together with the remaining, weaker negative couplings, they can almost completely cancel out all positive contributions.

Finally, at 350\,nm, no dominating parts can be clearly identified in the GCM. Positive and negative contributions 
cancel each other exactly. At even larger excitation wavelengths, the situation remains essentially unchanged, and the 
overall intensity is always zero. In summary, we have seen that the strongly positive ROA intensity under resonance conditions
originates from the NH\textsubscript{2} groups. These contributions quickly vanish when proceeding to larger excitation 
wavelengths, concomitantly reducing the total intensity of the corresponding normal mode.

\begin{figure}[H]
\begin{center}
\includegraphics[scale=0.18]{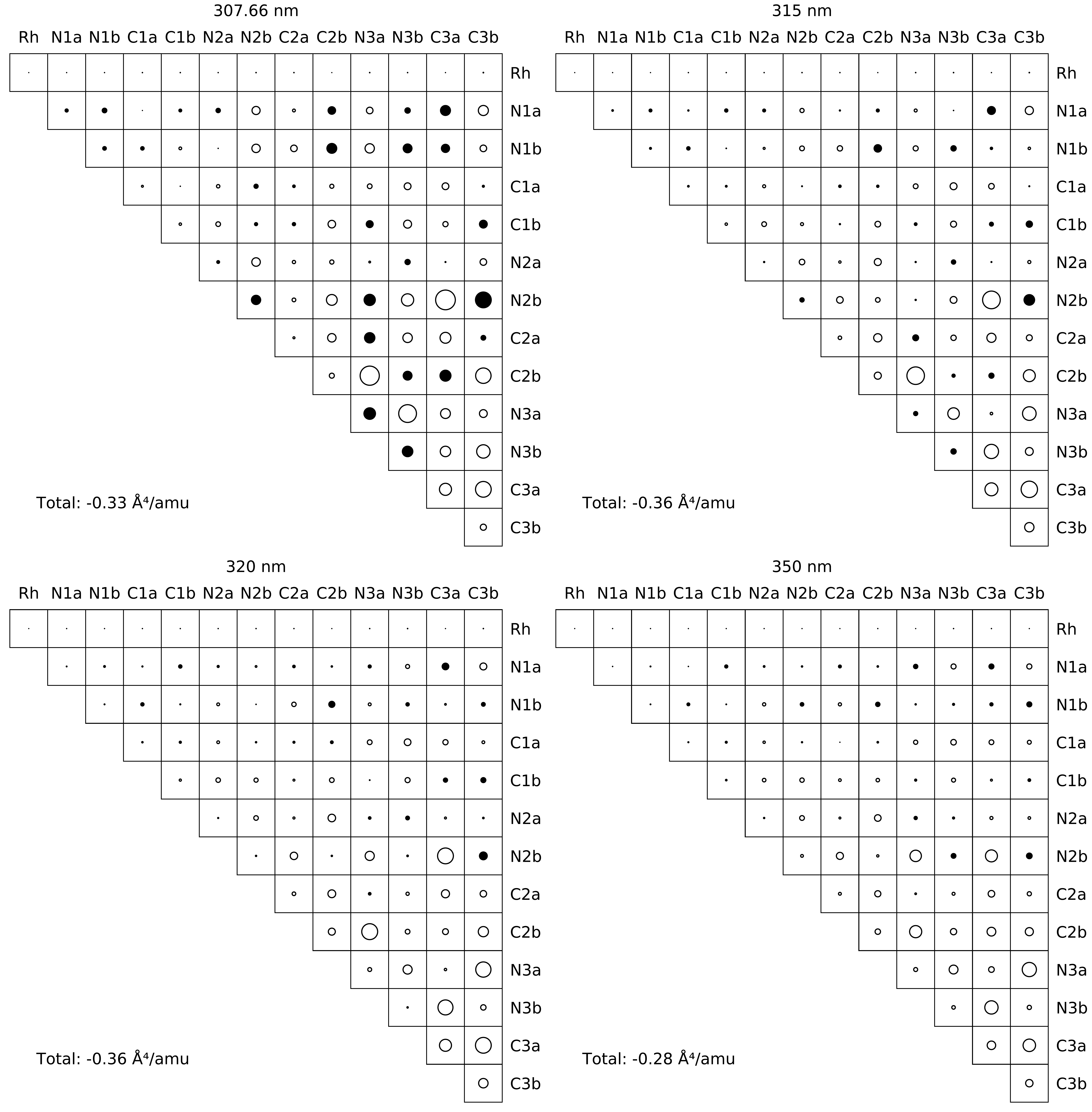}
\end{center}
\caption{\label{fig:gcm_1350}\small Group coupling matrices for the normal mode at 1350.4\,cm$^{-1}$ for four different 
excitation wavelengths (top left: 307.66\,nm; top right: 315\,nm; bottom left: 320\,nm; bottom right: 350\,nm). Positive 
contributions are represented by filled circles, while negative contributions are drawn as empty circles. The radius of 
the circle is proportional to the size of the corresponding GCM matrix element.}
\end{figure}

\begin{figure}[H]
\begin{center}
\includegraphics[scale=0.18]{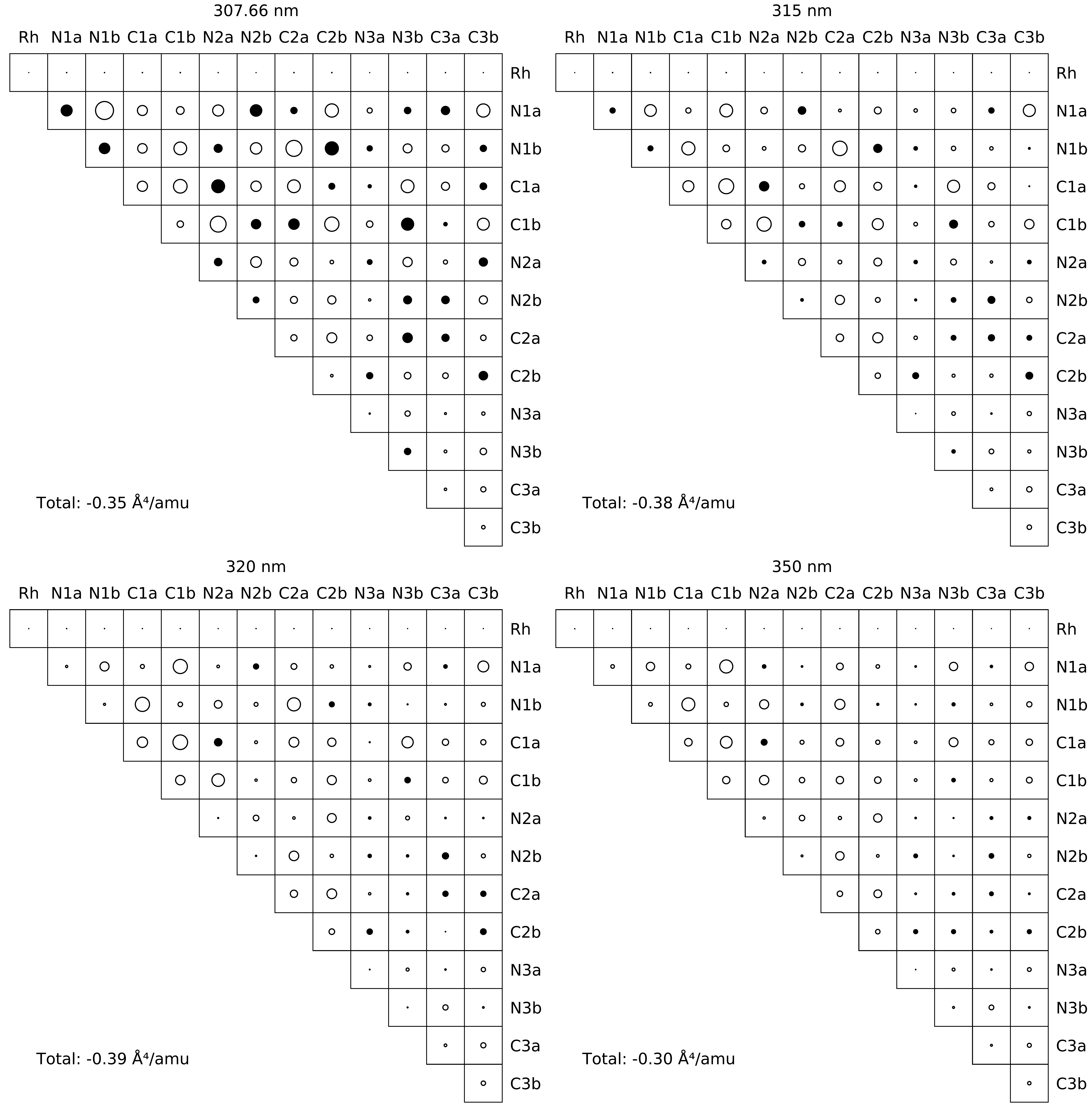}
\end{center}
\caption{\label{fig:gcm_1352}\small Group coupling matrices for the normal mode at 1352.5\,cm$^{-1}$ for four different 
excitation wavelengths (top left: 307.66\,nm; top right: 315\,nm; bottom left: 320\,nm; bottom right: 350\,nm). Positive 
contributions are represented by filled circles, while negative contributions are drawn as empty circles. The radius of 
the circle is proportional to the size of the corresponding GCM matrix element.}
\end{figure}

The negative band at about 1350\,cm$^{-1}$ results from three normal modes, at 1347.3\,cm$^{-1}$, at 1350.4\,cm$^{-1}$, 
and at 1352.5\,cm$^{-1}$. All three modes represent a twisting motion of the NH\textsubscript{2} and CH\textsubscript{2} 
groups (\textit{cf.}, Fig.~\ref{fig:normal_modes}). For the negative band around 1350\,cm$^{-1}$, the mode at 1347.3\,cm$^{-1}$ 
can be neglected to a first approximation, as its intensity is never larger than 0.02\,\AA{}$^4$\,amu$^{-1}$, while the other
two modes have intensities around 0.36\,\AA{}$^4$\,amu$^{-1}$ under resonance conditions. For the mode at 1350.4\,cm$^{-1}$, 
the origin of the total ROA intensity is difficult to identify (see Fig.~\ref{fig:gcm_1350}). Under resonance conditions, 
many couplings have roughly the same total strength. However, one can say that the most dominant contributions originate 
from the parts N2b\,--\,C3b, and couplings of these parts with other parts of the molecule. When going to 315\,nm, this 
general pattern remains unaffected, but many of the couplings are reduced in absolute value. However, the positive couplings 
appear to be reduced more than the negative couplings; hence, the total intensity even decreases to a certain degree. At 
320\,nm, almost all positive couplings between parts of N2b\,--\,C3b have vanished. At even larger wavelengths, the same 
general trend is observed: while the coupling pattern remains largely unaffected, the absolute intensities of all couplings 
diminish somewhat. Hence, the total intensity of this mode grows continuously weaker (albeit this decrease is by far not 
as strong as in the case of the band at around 520\,cm$^{-1}$). For the third mode of this band (at 1352.5\,cm$^{-1}$), 
the same general trend is valid (\textit{cf.}, Fig.~\ref{fig:gcm_1352}). The only difference is that under resonance 
conditions, most of the dominant coupling elements are among parts N1a\,--\,C1b.

Inspecting the GCMs shown in Figs.~\ref{fig:gcm_519}, \ref{fig:gcm_1350}, and \ref{fig:gcm_1352}, we also see that the 
rhodium atom never contributes significantly to the observed intensities, neither directly, nor through couplings with 
any of the other groups. We note that Humbert-Droz \textit{et al.}~have found the same result for the bands they analyzed 
in greater detail\cite{Humbert-Droz2014}, albeit only in the off-resonant case.

When analyzing the electronic excitations involved in the ROA spectra (see above), we find that electronic charge is moved 
from the metal center towards the NH\textsubscript{2} groups. It is therefore not too surprising that in the band around 
520\,cm$^{-1}$ (which changed significantly when approaching larger excitation wavelengths), most of the intensity under 
resonance conditions originates from exactly these NH\textsubscript{2} groups. We may hypothesize that any normal mode is 
particularly affected by resonance effects if the NH\textsubscript{2} groups are involved to a large extent. In order to 
elaborate on this, we calculated the amount to which these groups are involved in every normal mode. Furthermore, for 
every single normal mode, we identified the maximum as well as the minimum absolute intensity observed. The data for 
selected normal modes is shown in Table~\ref{tab:intensities}.

First of all, we see that the normal mode at 519.3\,cm$^{-1}$ features a rather large involvement of the NH\textsubscript{2} 
groups. This would also be expected, since this normal mode essentially represents a stretching vibration of all six 
Rh--N bonds. As already mentioned above, this mode features a large intensity under full resonance conditions (in fact, 
the highest intensity is even observed at a slightly shorter wavelength), and then collapses to a peak which is hardly 
identifiable at very large excitation wavelengths. Similarly, the almost constant intensity of the two normal modes at 
1350.4\,cm$^{-1}$ and 1352.5\,cm$^{-1}$ can be clearly seen. Since these normal modes represent a twisting motion of both
the CH\textsubscript{2} and the NH\textsubscript{2} groups, the latter groups have a much lower contribution to the total 
vibration. 

\begin{figure}[H]
\begin{center}
\includegraphics[scale=0.18]{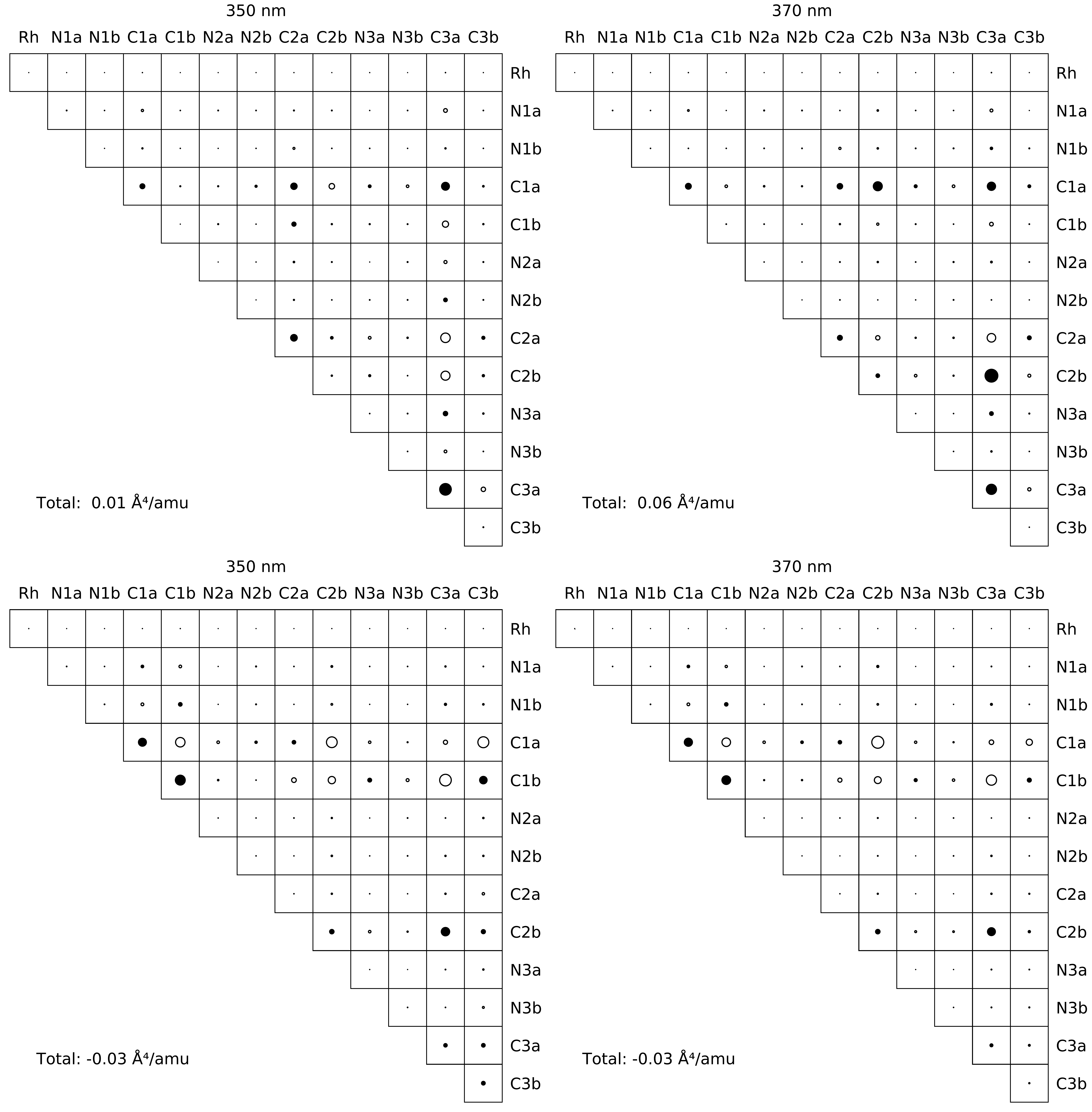}
\end{center}
\caption{\label{fig:gcm_1500}\small Group coupling matrices for the normal mode at 1494.2\,cm$^{-1}$ (top panel) and
the mode at 1496.6\,cm$^{-1}$ (bottom panel) for two different excitation wavelengths (left: 350\,nm; right: 370\,nm). 
Positive contributions are represented by filled circles, while negative contributions are drawn as empty circles. The 
radius of the circle is proportional to the size of the corresponding GCM matrix element.}
\end{figure}

Next, we study the two CH\textsubscript{2} scissoring vibrations at 1494.2\,cm$^{-1}$ and 1496.6\,cm$^{-1}$, respectively. 
Both normal modes have very similar wavenumbers and describe basically identical vibrations, albeit of different 
CH\textsubscript{2} groups. Naturally, the NH\textsubscript{2} groups do almost not participate in either of the two 
normal modes. However, the first of these normal modes (at 1494.2\,cm$^{-1}$) is strongly enhanced when moving to larger 
excitation wavelengths, reaching a maximum at 370\,nm, while the other normal mode is much less affected and follows 
exactly the opposite trend, featuring the highest intensity at the shortest excitation wavelength. This intriguing fact
can also be seen in the group coupling matrices of these two normal modes, shown in Fig.~\ref{fig:gcm_1500}. Note that 
the coupling matrix of the first normal mode changes significantly and in a qualitative fashion between 350\,nm and 370\,nm. 
For example, the coupling between the two CH\textsubscript{2} groups C1a and C2b (\textit{cf.}, Fig.~\ref{fig:gcm_partitioning}) 
is negative at 350\,nm, but it becomes strongly positive at 370\,nm. Likewise, the coupling between groups C2b and C3a 
changes from a negative contribution at 350\,nm to a positive contribution at 370\,nm. Since these contributions are also 
rather large in magnitude (in fact, they are the largest couplings across the entire GCM), the total intensity of this 
normal mode is significantly increased. In contrast to this, the corresponding GCMs for the second normal mode are almost 
identical to each other. Furthermore, this example clearly shows that the amount of NH\textsubscript{2} involvement in 
a given normal mode alone is not sufficient as a proxy for the prediction of how strongly this mode will be affected by 
resonance effects. Clearly, the first of these two modes is strongly affected by resonance effects, but the NH\textsubscript{2} 
groups are almost not involved in its vibration.

\begin{landscape}
\begin{table}[H]
\renewcommand{\baselinestretch}{1.0}
\renewcommand{\arraystretch}{1.0}
\caption{\label{tab:intensities}\small Wavenumbers $\omega$, maximum intensities $I_{\mathrm{max}}$ and corresponding 
excitation wavelengths $\lambda_{\mathrm{max}}$, minimum intensities $I_{\mathrm{min}}$ and corresponding excitation 
wavelengths $\lambda_{\mathrm{min}}$, the amount to which the NH\textsubscript{2} groups are involved and a description 
for selected normal modes.}
\begin{center}
\begin{tabular}{r r r r r r r l} \hline \hline
$\omega$ / cm$^{-1}$   & $I_{\mathrm{max}}$ / \AA{}$^4$\,amu$^{-1}$ & $\lambda_{\mathrm{max}}$ / nm & $I_{\mathrm{min}}$ / \AA{}$^4$\,amu$^{-1}$ & $\lambda_{\mathrm{min}}$ / nm & $I_{\mathrm{max}}$ / $I_{\mathrm{min}}$ & NH\textsubscript{2} Involvement & Description of Vibration \\
\hline 
 519.3	 & 3.299287	& 303	& 0.004048	& 700	   &  815.04	& 0.77	& Rh--N stretching \\
1350.4	 & 0.372716	& 315	& 0.064229	& 700	   &    5.80	& 0.39	& CH\textsubscript{2} twisting, NH\textsubscript{2} twisting \\
1352.5	 & 0.394029	& 320	& 0.06484	& 700	   &    6.08	& 0.38	& CH\textsubscript{2} twisting, NH\textsubscript{2} twisting \\
1494.2	 & 0.060442	& 370	& 4.2\,$\cdot\,10^{-5}$	& 303	   & 1439.10	& 0.10	& CH\textsubscript{2} scissoring \\
1496.6	 & 0.04483	& 290	& 0.011166	& 700	   &    4.01	& 0.09	& CH\textsubscript{2} scissoring \\
1639.4	 & 0.001732	& 303	& 1.1\,$\cdot\,10^{-5}$	& 700	   &  157.45	& 0.93	& NH\textsubscript{2} scissoring \\
1652.8	 & 0.172522	& 325	& 0.020486	& 307.66   &    8.42	& 0.88	& NH\textsubscript{2} scissoring \\
1653.4	 & 0.198994	& 290	& 0.014002	& 700	   &   14.21	& 0.88	& NH\textsubscript{2} scissoring \\
1658.2	 & 0.463895	& 290	& 0.056235	& 700	   &    8.25	& 0.88	& NH\textsubscript{2} scissoring \\
1664.6	 & 0.487597	& 303	& 0.0066	& 330	   &   73.88	& 0.91	& NH\textsubscript{2} scissoring \\
1665.2	 & 0.557156	& 303	& 0.025395	& 330	   &   21.94	& 0.90	& NH\textsubscript{2} scissoring \\
\hline
\hline
\end{tabular}
\renewcommand{\baselinestretch}{1.0}
\renewcommand{\arraystretch}{1.0}
\end{center}
\end{table}
\end{landscape}

As a final example, it is also instructive to examine the six normal modes representing NH\textsubscript{2} scissoring 
vibrations more closely. As we can see, in these normal modes there are almost only the NH\textsubscript{2} groups 
vibrating. As in the example before, these normal modes are very similar to each other in that they describe basically 
identical vibrations, and they feature similar wavenumbers. In this case, all six normal modes show the same qualitative 
trend for different excitation wavelengths: They all feature higher intensities at shorter wavelengths, and lower 
intensities at longer wavelengths. However, they are affected to different degrees. This is best seen when comparing the 
two modes at 1658.2\,cm$^{-1}$ and 1664.6\,cm$^{-1}$ with one another. While both share about the same maximum intensity,
their minimum intensities largely differ. While the first mode is only reduced by a factor of about eight, the intensity 
of the second mode is affected much more strongly, being reduced by more than a factor of 70.

In summary, we have seen that a complex electronic structure with several close-lying states leads to correspondingly 
complex (resonance) ROA spectra. While we have several tools at our disposal to analyze these spectra in great detail, 
it remains difficult to identify simple rules that allow one to predict whether a given normal mode will change a lot in 
intensity or not under resonance conditions.

\section{Conclusions}
\label{sec:conclusion}

We calculated the (resonance) ROA spectrum of [Rh(en)\textsubscript{3}]\textsuperscript{3+} at a range of different
wavelengths. These spectra were then examined, in particular with regard to resonance effects. Since the lowest two 
electronic excitations are almost equal in energy and also the third lowest excited state is rather close, there are 
always multiple states involved in these resonance effects. Therefore, the resulting spectra are never monosignate. 
Large differences can be observed between the spectrum at full-resonant wavelength and the spectra at off-resonance 
wavelengths. In many cases, the intensity of a given normal mode is enhanced significantly through these resonances, 
while other normal modes are almost unaffected or even de-enhanced (this being the consequence of the involvement of 
multiple electronically excited states). We analyzed two such modes in greater detail by means of the group coupling 
matrices proposed by Hug. In the case of the first normal mode, which was significantly enhanced under resonance condition, 
the intensity originated mainly in the NH\textsubscript{2} groups and the couplings between them. For larger excitation 
wavelengths, these contributions became significantly weaker, and concomitantly the intensity of this normal mode 
decreased. In the case of the second mode, the intensity of which stays almost constant across all excitation wavelengths, 
the ROA intensity is generated almost equally across all individual groups of the molecule, and no clear pattern can be 
established.

When analyzing the molecular orbitals of the lowest three electronic excitations, these excitations can all be interpreted 
as moving electronic charge from the metal center to the NH\textsubscript{2} groups. Therefore, many modes which involve 
NH\textsubscript{2} vibrations are strongly enhanced, while many other modes not involving these groups are not. However, 
there are exceptions to both cases. Therefore, it remains difficult to establish simple but still generally valid rules 
which would allow one to predict how strongly the intensity of a particular normal mode is affected by resonance effects.

\section*{Supporting Information}

Cartesian coordinates of [Rh(en)\textsubscript{3}]\textsuperscript{3+}, investigation of the influence of the damping
parameter $\Gamma$, verification of origin independence of the ROA spectra, an example input file, as well as the 
intensities of all normal modes as a function of the excitation wavelength.

\section*{Acknowledgments}

The author would like to thank Prof.~M.~Reiher (ETH Zurich) for continuous support and many inspiring discussions. 

\providecommand{\latin}[1]{#1}
\makeatletter
\providecommand{\doi}
  {\begingroup\let\do\@makeother\dospecials
  \catcode`\{=1 \catcode`\}=2\doi@aux}
\providecommand{\doi@aux}[1]{\endgroup\texttt{#1}}
\makeatother
\providecommand*\mcitethebibliography{\thebibliography}
\csname @ifundefined\endcsname{endmcitethebibliography}
  {\let\endmcitethebibliography\endthebibliography}{}

\newpage

\begin{center}
 \textbf{TOC Graphic}
\end{center}

\begin{figure}[H]
\begin{center}
\includegraphics[scale=1.0]{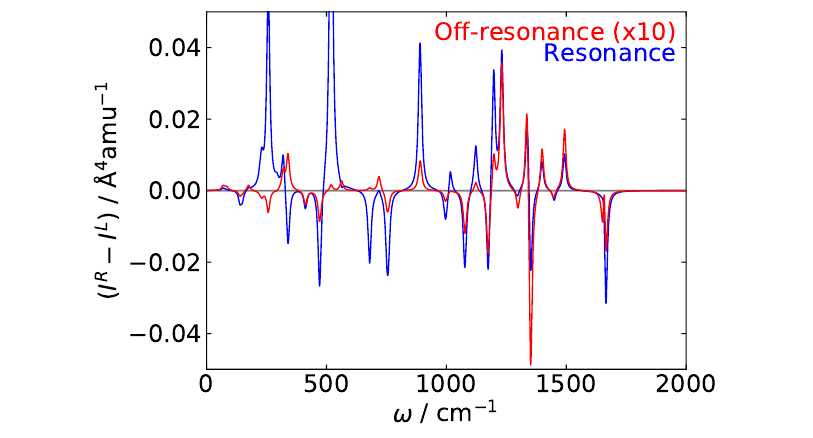}
\end{center}
\end{figure}

\end{document}


\begin{center}

{\LARGE\bf
 Resonance Effects in the Raman Optical Activity Spectrum of [Rh(en)\textsubscript{3}]\textsuperscript{3+}
}

\vspace{1cm}

{\large
Thomas Weymuth$^{a,}$\footnote{Corresponding author; e-mail: thomas.weymuth@phys.chem.ethz.ch; ORCID: 0000-0001-7102-7022}
}\\[4ex]

$^{a}$ Laboratory for Physical Chemistry, ETH Zurich, \\
8093 Zurich, Switzerland

September 30, 2019

\vspace{.41cm}

\end{center}

\begin{center}
\textbf{Supporting Information}
\end{center}

\newpage

\section{Cartesian Coordinates of [Rh(en)\textsubscript{3}]\textsuperscript{3+}}

Below we give the Cartesian coordinates of [Rh(en)\textsubscript{3}]\textsuperscript{3+} as obtained from the 
structure optimization. All values are given in \AA{}.

\begin{verbatim}
C                    -2.56007083    -1.35708902    -0.54082819
H                    -3.10328879    -2.24869940    -0.86238831
H                    -3.08660763    -0.49282315    -0.95265826
C                    -2.47421399    -1.28021359     0.95943603
H                    -3.46617474    -1.19818162     1.40985356
H                    -2.00345265    -2.17345255     1.37692725
N                    -1.16889899    -1.35936049    -1.09006688
H                    -1.21286788    -1.16089489    -2.08908158
H                    -0.80398432    -2.31021408    -1.04209812
N                    -1.63162263    -0.09834763     1.31900179
H                    -1.37112102    -0.16978020     2.30207886
H                    -2.21364735     0.73774283     1.27663904
Rh                   -0.00002018     0.00010660     0.00233207
C                    -0.02679481     2.88428537    -0.60574077
C                     2.35191497    -1.42093764    -1.06582865
H                    -0.58752840     3.79240567    -0.83894857
H                     3.35099977    -1.38928934    -1.50663534
H                     0.88253316     2.90273817    -1.21114681
H                     1.80857443    -2.23169584    -1.55717426
C                     0.29510901     2.80224385     0.86182796
C                     2.41748095    -1.63037140     0.42278229
H                     0.93417380     3.63108091     1.17516908
H                     2.87727091    -2.59042055     0.66869121
H                    -0.61140381     2.84628607     1.47013288
H                     3.01413515    -0.85371141     0.90728810
N                    -0.81152322     1.66769055    -0.98163326
N                     1.61872404    -0.14367645    -1.32703344
H                     1.35244579    -0.11627064    -2.31077538
H                    -0.82072619     1.58682826    -1.99789514
H                     2.27364722     0.63115483    -1.22583125
H                    -1.78870042     1.82488763    -0.73660797
N                     0.96576365     1.49089727     1.12047395
N                     1.02648486    -1.55627954     0.96667280
H                     0.97308157     1.32058012     2.12565099
H                     1.07697248    -1.45807464     1.98016955
H                     1.95183719     1.57898160     0.87656334
H                     0.57692005    -2.46074498     0.82712764
\end{verbatim}

\section{Damping Factor}

\begin{figure}[H]
\begin{center}
\includegraphics[scale=0.9]{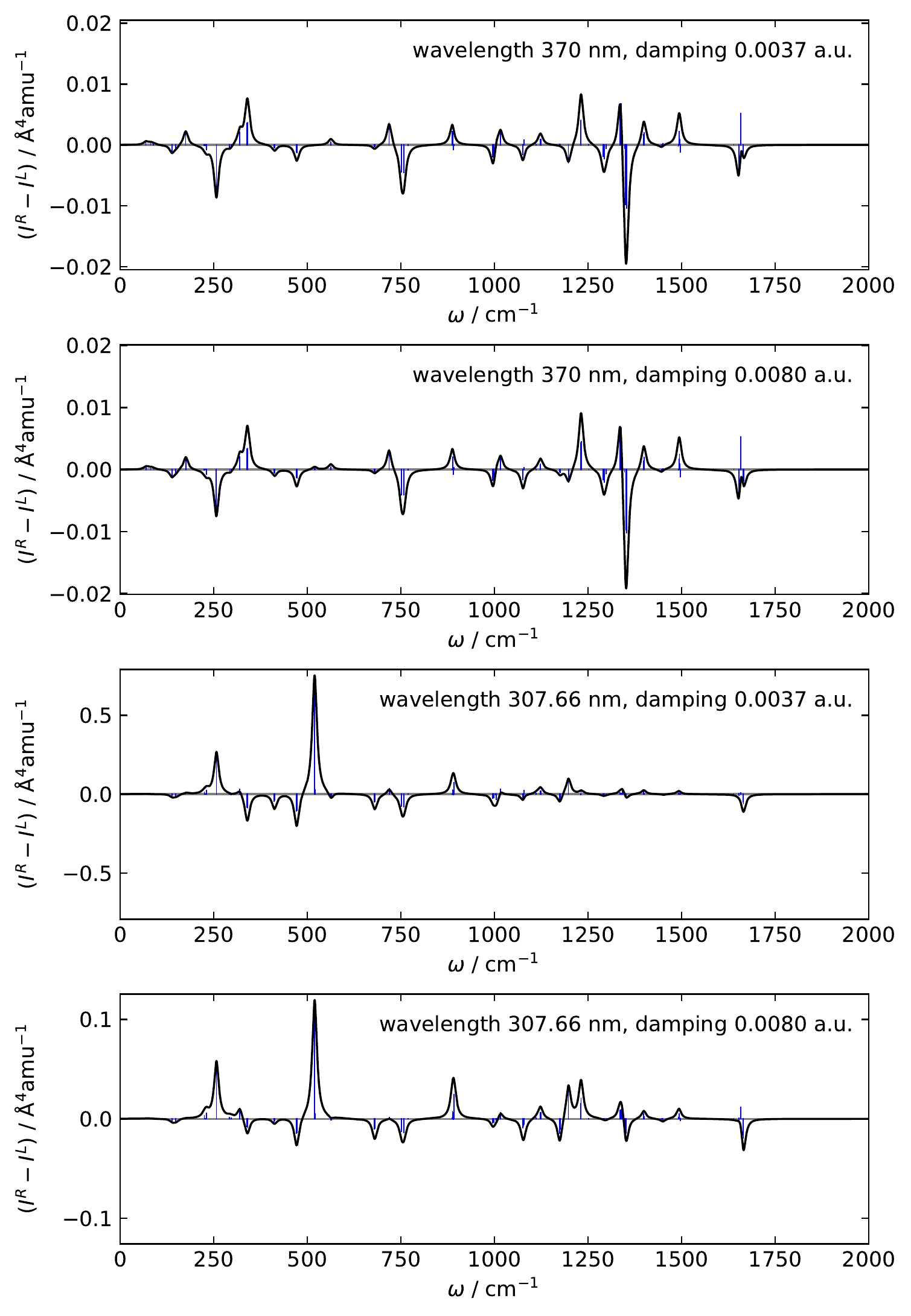}
\end{center}
\caption{\label{fig:damping}\small ROA spectra of [Rh(en)\textsubscript{3}]\textsuperscript{3+} for two different excitation 
wavelengths and two different damping values. A wavelength of 307.66\,nm yields full-resonance spectra.
Note that the individual spectra span different intensity ranges.}
\end{figure}

In order to investigate the influence of the damping value, we calculated the ROA spectrum of [Rh(en)\textsubscript{3}]\textsuperscript{3+}
for two different damping values. In the first case, we adopted a rather small damping value of 0.0037\,a.u.; in the
second case, a larger value of 0.008\,a.u.~was used. For both damping values, we calculated the spectrum with an excitation 
wavelength of 307.66\,nm, \textit{i.e.},under full-resonance conditions, and with an excitation wavelength of 370\,nm.
The four resulting spectra are shown in Fig.~\ref{fig:damping}. As one can see, when an excitation wavelength of 370\,nm
is adopted, the two different damping values lead to virtually identical spectra, and therefore have no influence (\textit{cf.},
the top two spectra in Fig.~\ref{fig:damping}). Under full-resonance conditions, however, the damping value has a rather
large influence on the overall spectral intensity. With larger damping value, the overall intensity is greatly reduced
(up to a factor of about five). However, the relative intensities as well as the overall band shape is much less affected
by changes in the damping value. We therefore conclude that the exact numerical value adopted for the damping value is
not of primary importance as long as one and the same value is used throughout our study. This will suffice to make the 
individual spectra comparable to each other.

\section{Origin Dependence}

\begin{figure}[H]
\begin{center}
\includegraphics[scale=0.9]{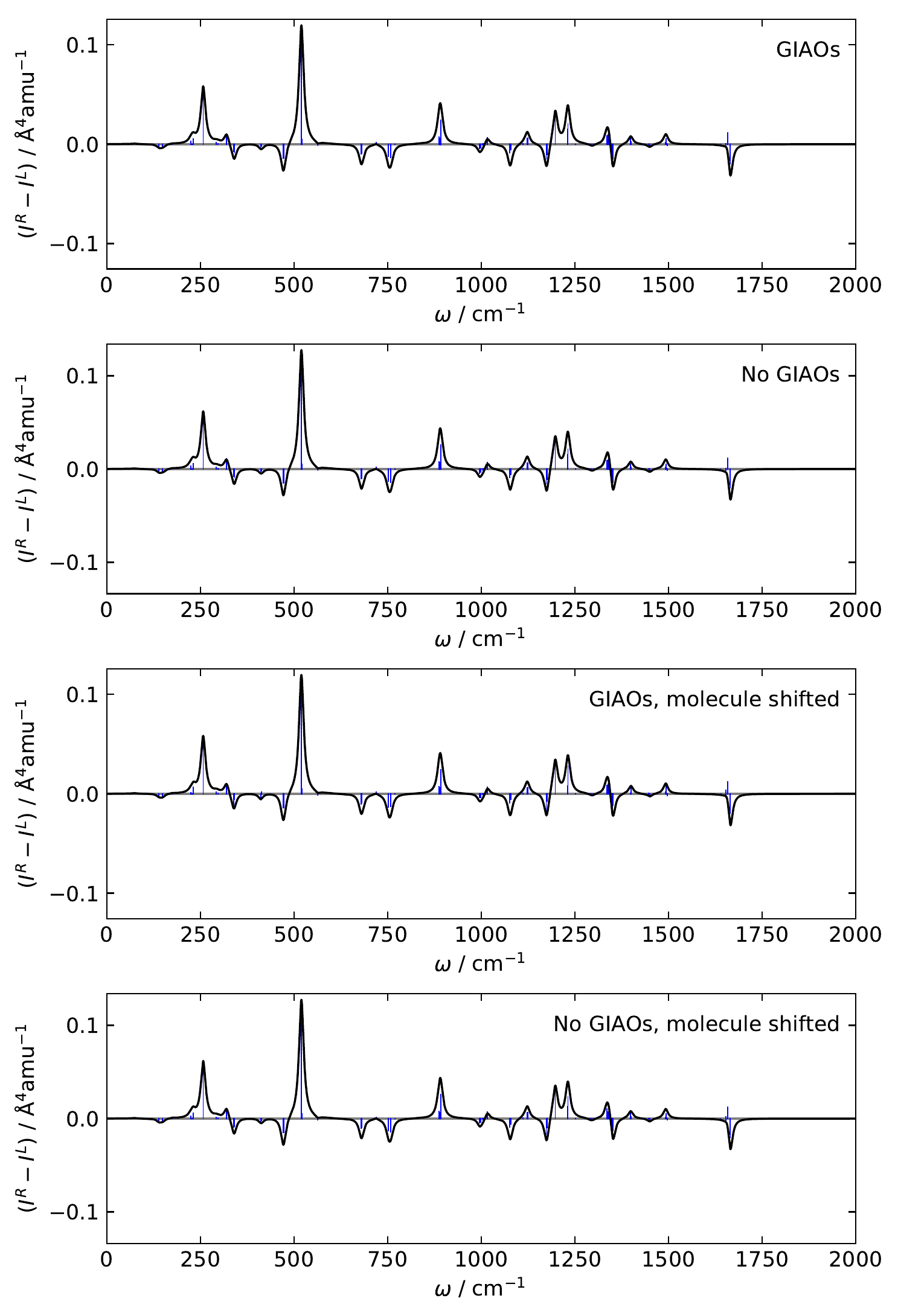}
\end{center}
\caption{\label{fig:origin_independence}\small ROA spectra of [Rh(en)\textsubscript{3}]\textsuperscript{3+} calculated either with or 
without gauge-including atomic orbitals. In the upper two spectra, the molecule is placed very close to the origin,
while in the lower two spectra, the molecule is shifted by 10\,\AA{} in all three spatial directions.}
\end{figure}

In order to check whether the methodology followed in this work yields ROA spectra which are origin independent, we shifted
the molecule by 10\,\AA{} in each of the three spatial directions and recalculated the spectrum under full-resonance
conditions. As can be seen by comparing the first and third spectrum in Fig.~\ref{fig:origin_independence}, the spectrum
does literally not change. We therefore conclude that our spectra are not dependent on the choice of origin.

As a further test, we also calculated the ROA spectrum under full-resonance conditions without gauge-including atomic
orbitals (GIAOs), both with the molecule at its original location and with the molecule shifted by 10\,\AA{} in all three
spatial directions. As can be seen from Fig.~\ref{fig:origin_independence}, in our case the use of GIAOs apparently has 
no observable effects on the ROA spectrum.

\section{Example Input File}

Below we give an example \textsc{NWChem} input file as used in this work to calculate the ROA spectra. The nuclear coordinates
have been omitted in this example; in the actual calculations, the distorted structures as generated by \textsc{MoViPac}
have been inserted.

\begin{verbatim}
echo

title "RROA calculation of [Rh(en3)]3+ in vacuum"

memory 2048 mb

start rh_en

geometry units au nocenter noautosym nucleus point
symmetry c1
...
end

basis "ao basis" spherical
 * library def2-TZVP
end

ecp
 Rh library def2-ECP
end

scf
 singlet
 rhf
 direct
end

charge 3

dft
 grid xfine nodisk
 tolerances tight
 convergence energy 1d-8
 convergence density 1d-7
 maxiter 50
 xc pbe0 
 disp vdw 3
 direct
 noio
end

set prop:newaoresp 1
set aor:roadata T
set cphf:maxsub 30
set cphf:maxiter 50

property
  response 1 0.14809645
  bdtensor
  convergence 1e-4
  damping 0.0080
  giao
end

task dft property
task dft gradient
\end{verbatim}

\section{Intensities of Normal Modes}

\begin{figure}[H]
\begin{center}
\includegraphics[scale=0.33]{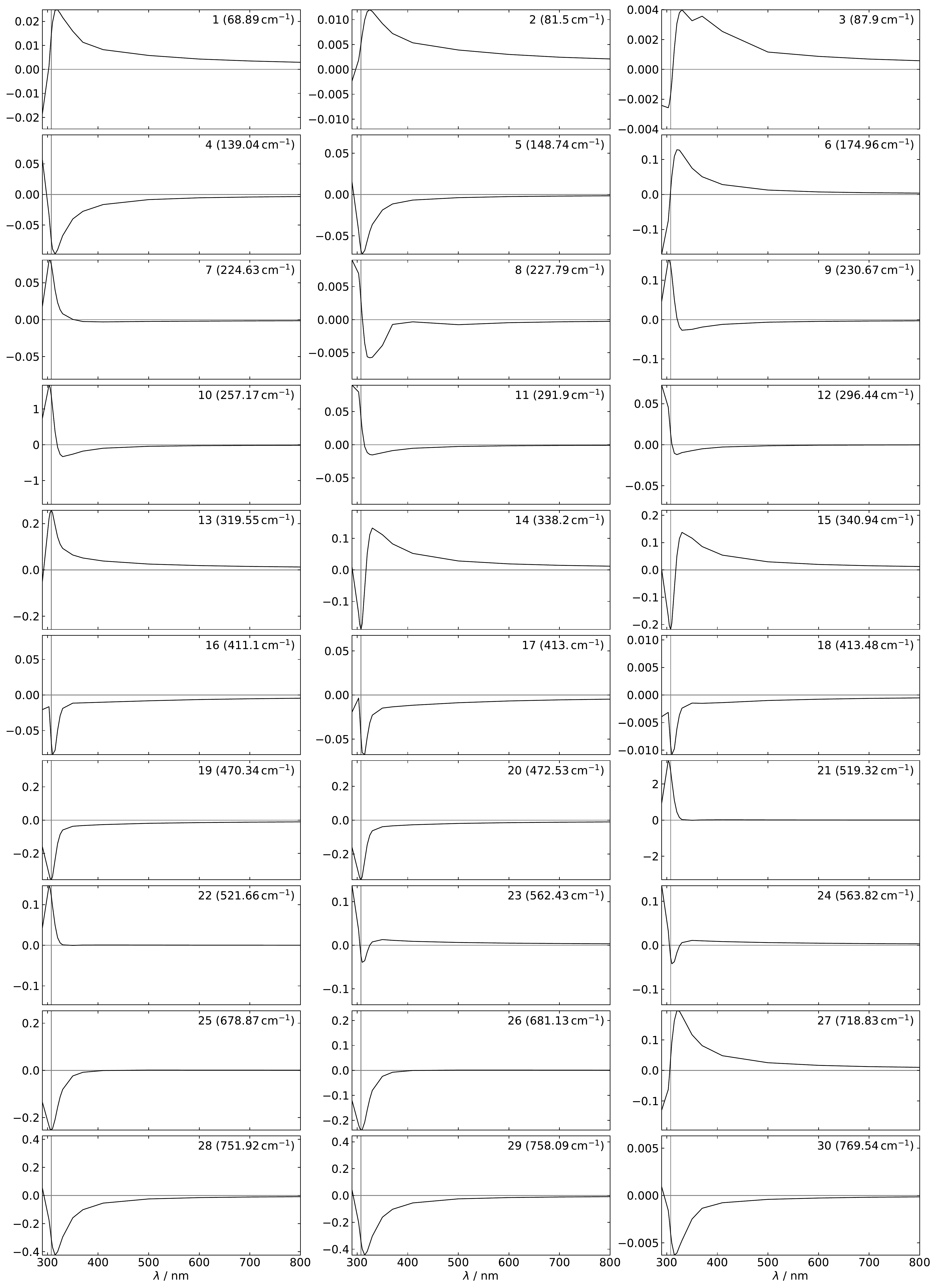}
\end{center}
\caption{\label{fig:intensities1}\small Intensities of normal modes 1 to 30 as a function of excitation wavelength. All intensities
are given in \AA{}$^4$\,/\,amu$^{-1}$.  The vertical line highlights the full-resonance wavelength.}
\end{figure}

\begin{figure}[H]
\begin{center}
\includegraphics[scale=0.33]{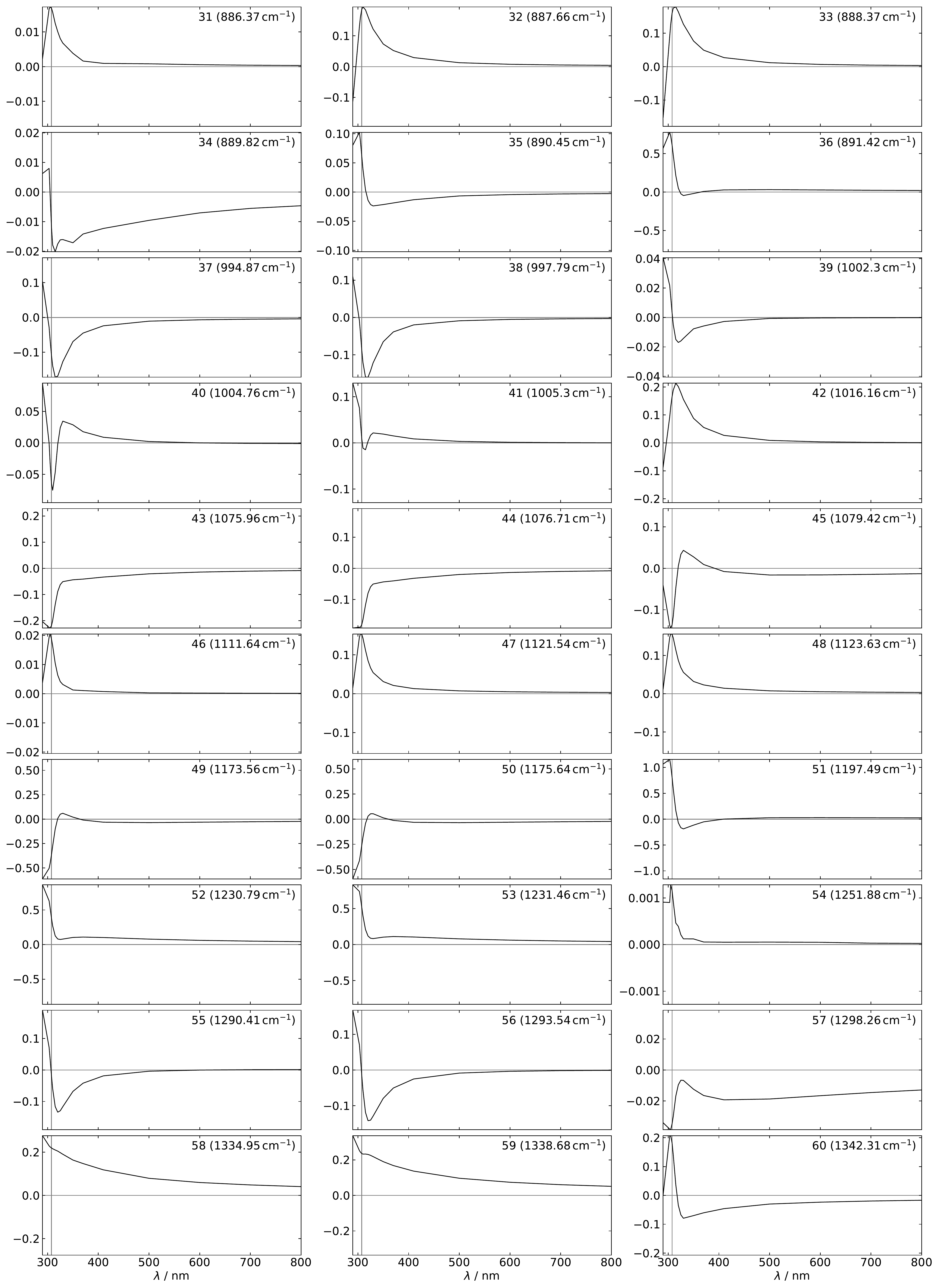}
\end{center}
\caption{\label{fig:intensities2}\small Intensities of normal modes 31 to 60 as a function of excitation wavelength. All intensities
are given in \AA{}$^4$\,/\,amu$^{-1}$.  The vertical line highlights the full-resonance wavelength.}
\end{figure}

\begin{figure}[H]
\begin{center}
\includegraphics[scale=0.33]{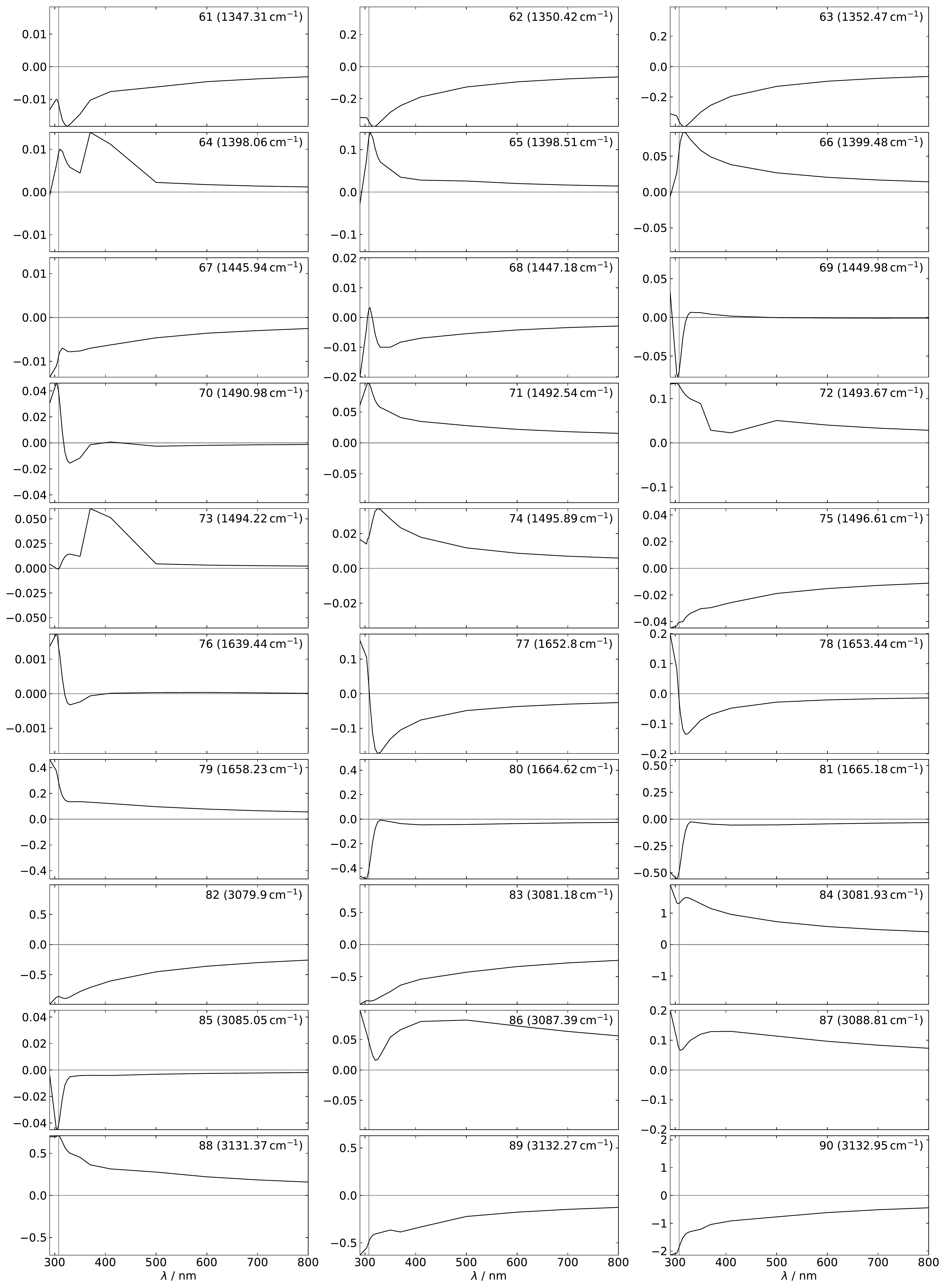}
\end{center}
\caption{\label{fig:intensities3}\small Intensities of normal modes 61 to 90 as a function of excitation wavelength. All intensities
are given in \AA{}$^4$\,/\,amu$^{-1}$.  The vertical line highlights the full-resonance wavelength.}
\end{figure}

\begin{figure}[H]
\begin{center}
\includegraphics[scale=0.33]{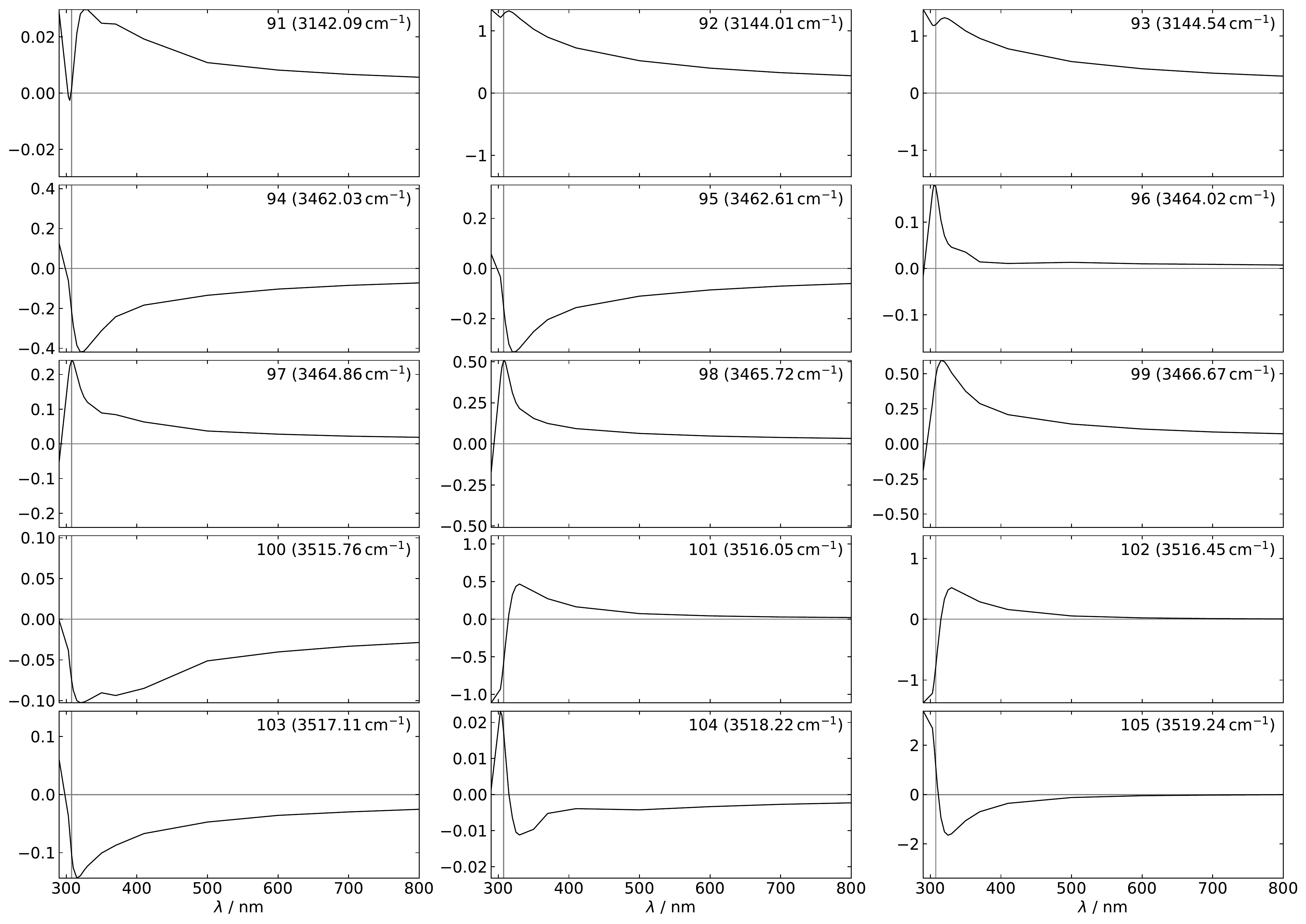}
\end{center}
\caption{\label{fig:intensities4}\small Intensities of normal modes 91 to 105 as a function of excitation wavelength. All intensities
are given in \AA{}$^4$\,/\,amu$^{-1}$. The vertical line highlights the full-resonance wavelength.}
\end{figure}